\documentclass[a4paper,11pt]{article}
\pdfoutput=1 

\usepackage{jheppub} 

\usepackage[T1]{fontenc}
\usepackage[italian,english]{babel}
\usepackage{hyperref}
\usepackage{ifpdf}
\usepackage{subfigure}
\usepackage{dutchcal}
\usepackage{amssymb}
\usepackage{amsfonts}
\usepackage{epsf}
\usepackage{rotating}
\usepackage{graphicx}
\usepackage{amsmath}
\usepackage{fancyhdr}
\usepackage{lineno}
\usepackage{babel}
\usepackage{graphics}
\usepackage{pstricks}
\usepackage{color}
\usepackage{multirow}
\usepackage{hhline}
\newcommand{\nn}{\nonumber}
\newcommand{\lsim}{\mathrel{\mathop{\kern 0pt \rlap
  {\raise.2ex\hbox{$<$}}}
  \lower.9ex\hbox{\kern-.190em $\sim$}}}
\newcommand{\gsim}{\mathrel{\mathop{\kern 0pt \rlap
  {\raise.2ex\hbox{$>$}}}
  \lower.9ex\hbox{\kern-.190em $\sim$}}}

\newcommand{\be}{\begin{equation}}
\newcommand{\ee}{\end{equation}}
\newcommand{\bea}{\begin{eqnarray}}
\newcommand{\eea}{\end{eqnarray}}

\def\etmiss{\not\!\!{E_T}}
\def\ptmiss{\not\!\!{p_T}}

\newcommand{\pd}{\partial}
\newcommand{\beq}{\begin{equation}}
\newcommand{\eeq}{\end{equation}}

\newcommand{\beqa}{\begin{eqnarray}}
\newcommand{\eeqa}{\end{eqnarray}}

\let\a   = \alpha     \let\b = \beta        \let\d = \delta

\let\s   = \sigma

\title{\boldmath Bounds on the Conformal Scale of a Minimally Coupled Dilaton and Multi-Leptonic Signatures at the 
LHC}

\author[a]{Priyotosh Bandyopadhyay}

\author[a,b]{Claudio Corian\`o}
\author[a]{Antonio Costantini}
\author[c,d]{Luigi Delle Rose}

\affiliation[a]{Dipartimento di Matematica e Fisica "Ennio De Giorgi", \\ Universit\`a del Salento and INFN-Lecce, \\ Via Arnesano, 73100 Lecce, Italy}
\affiliation[b]{STAG Research Centre and Mathematical Sciences,\\ University of Southampton, Southampton SO17 1BJ, UK}
\affiliation[c]{School of Physics and Astronomy, University of Southampton, \\Highfield, Southampton SO17 1BJ, UK}
\affiliation[d]{Dept. of Particle Physics, Rutherford Appleton Laboratory, \\Chilton, Didcot, OX11 0QX, UK}

\emailAdd{priyotosh.bandyopadhyay@le.infn.it}
\emailAdd{claudio.coriano@le.infn.it}
\emailAdd{antonio.costantini@le.infn.it}
\emailAdd{l.delle-rose@soton.ac.uk}

\abstract{ We explore the potential for the discovery of a dilaton $\mathcal{O}(200-500)$ GeV in a classical scale/conformal invariant extension of the Standard Model by investigating the size of the corresponding breaking scale $\Lambda$ at the LHC, extending a previous analysis. In particular, we address the recent bounds on $\Lambda$ derived from Higgs boson searches.  We investigate if such a  dilaton can be produced via 
gluon-gluon fusion, presenting rates for its decay either into a pair of Higgs bosons or into two heavy gauge bosons, which can give rise to multi-leptonic final states. A detailed analysis via PYTHIA-FastJet has been carried out of the dominant Standard Model backgrounds, at a centre of mass energy of 14 TeV. We show that early data of $\sim 20$ fb$^{-1}$ can certainly probe the region of parameter space where such a dilaton is allowed. A conformal scale of 5 TeV is allowed by the current data, for almost all values of the dilaton mass investigated.}

\begin{document}

\maketitle
\flushbottom

\section{Introduction}

 An important feature of the electroweak sector of the Standard Model (SM) is its approximate scale invariance which holds if the quadratic terms of the Higgs potential are absent. 
These terms are obviously necessary in order for the theory to be in a spontaneously broken phase with a vacuum expectation value (vev) $v$ which is fixed by the experiments. \\
The issue of incorporating a mechanism of spontaneous symmetry breaking of a gauge symmetry while preserving the scale invariance of the Lagrangian is a subtle one, 
which naturally brings to the conclusion that the breaking of this symmetry has to be dynamical, with the inclusion of a dilaton field. In this case the mass of the dilaton should be attributed to a specific symmetry-breaking potential, probably of non-perturbative origin. A dilaton, in this case, is likely to be a composite \cite{CDS} state, with a conjectured behaviour which can be partly discussed using the conformal anomaly action.  

The absence of any dimensionful constant 
in a tree level Lagrangian is, in fact, a necessary condition in order to guarantee the scale invariance of the theory.  This is also the framework that we will consider, which is based on the requirement of {\em classical} scale invariance. A stricter condition, for instance, lays in the (stronger) requirement of quantum scale invariance, with correlators which, in some cases, are completely fixed by the symmetry and incorporate the anomaly \cite{OP, EO, BMS1,BMS2,BMS3}.  
In the class of theories that we consider, the invariance of the Lagrangian under special conformal transformations are automatically fulfilled by the condition of scale invariance. For this reason we will 
refer to the breaking of such symmetry as to a conformal breaking.\\
 Approaching a scale invariant theory from a non scale-invariant one requires all the dimensionful couplings of the model to be turned into dynamical fields, with a compensator ($\Sigma(x)$) which is rendered dynamical by the addition of a scalar kinetic term. It is then natural to couple such a field both to the anomaly and to the explicit (mass-dependent) extra terms which appear in the classical trace of the stress-energy tensor. \\
 The inclusion of an extra $\Sigma$-dependent potential in the scalar sector of the new theory is needed in order to break the conformal symmetry at the TeV scale, with a dilaton mass which remains, essentially, a free parameter.   We just mention that for a classically scale invariant extension of the SM Lagrangian, the choice of the scalar potential has to be appropriate, in order to support a spontaneously broken phase of the theory, such as the electroweak phase \cite{CDS}. For such a reason, the two mechanisms of electroweak and scale breaking have to be directly related, with the electroweak scale $v$ and the conformal breaking scale $\Lambda$ linked by a simple expression. At the same time, the invariance of the action under a change induced by a constant shift of the potential, which remains unobservable in a non scale-invariant theory, 
becomes observable and affects the vacuum energy of the model and its stability.  \\
The goal of our work is to elaborate on a former theoretical analysis \cite{CDS} of dilaton interactions, by discussing the signatures and the phenomenological bounds on a possible state of this type at the LHC, using the current experimental constraints. Some of the studies carried so far address a state of geometrical origin ({\em the radion}) \cite{GRW}, which shares several of the properties of a (pseudo) Nambu-Goldstone mode of a broken conformal symmetry, except, obviously, its geometric origin and its possible compositeness. Other applications are in inflaton physics (see for instance \cite{AR}). \\ 
The production and decay mechanisms of a dilaton, either as a fundamental or a composite state, are quite similar to those of the Higgs field, except for the presence of a suppression related to a conformal scale ($\Lambda$) and of a direct contribution derived from the conformal anomaly. As we are going to show, the latter causes an enhancement of the dilaton decay modes into massless states, which is maximized if its coupling $\xi$ is conformal.

\subsection{The role of compositeness} 
In the phenomenological study that we present below we do not consider possible modifications of the production and decay rates of this particle typical of the dynamics of a bound state, if a dilaton is such. 
This point would require a separate study that will be addressed elsewhere. We just mention that there are significant indications from the study of conformal anomaly actions \cite{CDS,CCDS} both in ordinary and in supersymmetric theories, that the conformal anomaly manifests with the appearance of anomaly poles in specific channels. These interpolate with the dilatation current \cite{CDS}, similarly to the behaviour manifested by an axial-vector current in $AVV$ diagrams.
The exchange of these massless poles are therefore the natural signature of anomalies in general, being them either chiral or conformal  \cite{ACD0}. 
Concerning the conformal ones, these analyses have been fully worked out in perturbation theory in a certain class of correlators ($TVV$ diagrams) \cite{Giannotti:2008cv,Armillis:2009pq}, starting from QED. We have included one  
section (section \ref{non0xi}) where we briefly address these points, in view of some recent developements and prospects for future studies. In this respect, the analysis that we present should be amended with the inclusion of corrections coming from a possible wave function of the dilaton in the production/decay processes involving such a state. These possible developments require  specific assumptions which we are not going  to discuss in great detail in the current study but on which we will briefly comment prior to our conclusions.

Our work is organised as follows. In order to make our discussion self-contained, we will briefly review the salient features of dilaton interactions in section \ref{revv}. Afterwards we will turn to a numerical analysis of the possible final states which could be a direct signature of the exchange of a dilaton at the LHC.\\
The phenomenological study will start with a discussion of the decay modes of the dilaton in section~\ref{decays}, followed in section~\ref{prod} by an analysis of its dominant production modes at the LHC. 
These are characterised by a significant presence of leptons and missing transverse energy/momentum 
in the final state that we will quantify. 
These studies will allow us to present some bounds on the conformal scale $\Lambda$, and to identify some phenomenological channels for its possible experimental study, improving on a previous exclusion limit ($\sim 1$ TeV) \cite{wisc,wisc1}. In section~\ref{colsim} we present a PYTHIA based analysis of the dominant SM backgrounds with multi-lepton final states. Our perspectives for further analysis of dilaton production 
and decay, with the inclusion of corrections due to a possible composite nature of this state, are briefly discussed in section \ref{non0xi}, followed by our conclusions in section~\ref{concl}.

\section{ Classical scale invariant extensions of the Standard Model and dilaton interactions}
\label{revv}
A scale invariant extension of the SM, at tree level, can be trivially obtained by promoting all the 
dimensionful couplings in the scalar potential, which now includes quartic and quadratic Higgs terms,  to dynamical fields. The new field ($\Sigma(x)=\Lambda e^{\rho(x)/\Lambda}$) is accompanied by a conformal scale ($\Lambda$) and introduces a dilaton field $\rho(x)$, as a fluctuation around the vev of $\Sigma(x)$
 \beq
 \Sigma(x)= \Lambda + \rho(x) + O(\rho^2), \qquad \qquad \langle \Sigma(x) \rangle=\Lambda, \qquad \qquad \langle \rho(x) \rangle =0.
 \eeq
The inclusion of $\rho$, via an exponential, provides a nonlinear realization of the dilatation symmetry. 
 In this section we will briefly review the structure of the coupling of a dilaton field to the matter of the SM. 
 
\begin{figure}[t]
\begin{center}
\includegraphics[scale=1.2]{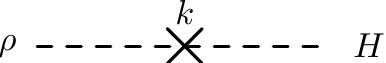}
\end{center}
\caption{Bilinear dilaton/Higgs vertex at tree level from the term of improvement.  }
\label{dilatonpole}
\end{figure}

 The leading interactions of the dilaton with the SM fields are obtained through the divergence of the dilatation current. This corresponds to the trace of the energy-momentum tensor $T^\mu_{\mu \, SM}$ computed on the SM fields
 
\bea\label{tmunu}
\mathcal L_{int} = -\frac{1}{\Lambda}\rho T^\mu_{\mu\,SM}.
\eea
The expression of the energy-momentum tensor can be easily derived by embedding the 
SM Lagrangian on the background metric $g_{\mu\nu}$ 
\beq S = S_{SM} + S_{I}=  \int d^4 x
\sqrt{-g}\mathcal{L}_{SM} + \xi \int d^4 x \sqrt{-g}\, R \, \mathcal H^\dag \mathcal H      \, ,
\eeq
where $ H$ is the Higgs doublet and $R$ the scalar curvature of the same metric, and then defining 
\beq  T_{\mu\nu}(x)  = \frac{2}{\sqrt{-g(x)}}\frac{\d [S_{SM} + S_I ]}{\d g^{\mu\nu}(x)},
\eeq
or, in terms of the SM Lagrangian, as
\beq \label{TEI spaziocurvo}
\frac{1}{2} \sqrt{-g} T_{\mu\nu}{\equiv} \frac{\pd(\sqrt{-g}\mathcal{L})}
{\pd g^{\mu\nu}} - \frac{\pd}{\pd x^\s}\frac{\pd(\sqrt{-g}\mathcal{L})}{\pd(\pd_\s g^{\mu\nu})}\, .
\eeq
The complete expression of the energy-momentum tensor can be found in \cite{Coriano:2011zk}.
$S_{I}$ is responsible for generating a term of improvement $(I)$, which induces a mixing between the Higgs and the dilaton after spontaneous symmetry breaking. 
As usual, we parameterize
the vacuum $\mathcal H_0$ in the scalar sector in terms of the electroweak vev $v$ as
\beq \label{VEVHiggs}
\mathcal H_0 =
\left(\begin{array}{c} 0 \\ \frac{v}{\sqrt{2}} \end{array}\right)
\eeq
and we expand the Higgs doublet in terms of the physical Higgs boson $H$ and the two Goldstone bosons $\phi^{+}$, $\phi$ as
\bea
\mathcal H = \left(\begin{array}{c} -i \phi^{+} \\ \frac{1}{\sqrt{2}}(v + H + i \phi) \end{array}\right),
\eea
obtaining from the term of improvement of the stress-energy tensor the expression
\beq
T^I_{\mu\nu} = - 2 \xi \bigg[ \partial_{\mu} \partial_{\nu} - \eta_{\mu\nu} \, \Box \bigg] \mathcal H^\dag \mathcal H = - 2 \xi \bigg[ \partial_{\mu} \partial_{\nu} - \eta_{\mu\nu} \, \Box \bigg] \bigg( \frac{H^2}{2} + \frac{\phi^2}{2} + \phi^{+}\phi^{-} + v \, H \bigg),
\eeq
which is responsible for a bilinear vertex shown in Fig. \ref{dilatonpole}
\beq
V_{I,\, \rho H }(k)
= - \frac{i}{\Lambda} \frac{12\,\xi \, s_w M_W}{e} \, k^2. \nn
\eeq
 The trace takes contribution from the massive fields, the fermions and the electroweak gauge bosons, and from the conformal anomaly (also dubbed trace-anomaly) in the massless gauge boson sector, through the $\beta$ functions of the corresponding coupling constants. In most of our numerical analysis we will consider a dilaton which is minimally coupled to the trace of the stress-energy tensor $(\xi=0)$, but we will release this constraint 
 in the final part of our work when we are going to briefly investigate the dependence of the decay rates on $\xi$. A general analysis of the steps involved in the derivation of the two mass eigenstates for the physical Higgs and the dilaton can be found in \cite{GRW}.  
  In a phenomenological context is expected that both for a fundamental or for a composed dilaton the leading interaction with the fields of the SM should be characterised by $T^\mu_{\mu\,SM}$.  \\
 The separation between the anomalous and the explicit mass-related terms in the expression of the correlators responsible of the conformal anomaly  can be directly verified in perturbation theory, in the computation of basic correlators with one insertion of the stress energy tensor \cite{Armillis:2009pq, CDS}. As pointed out in \cite{CDS}, one can check that in a 
mass-independent renormalization scheme, such as Dimensional Regularization with minimal subtraction, this separation holds. By tracing these correlators one derives an anomalous Ward identity of the form 
\beq
\Gamma^{\alpha\beta}(z,x,y) 
\equiv \eta_{\mu\nu} \left\langle T^{\mu\nu}(z) V^{\alpha}(x) V'^{\beta}(y) \right\rangle 
= \frac{\delta^2 \mathcal A(z)}{\delta A_{\alpha}(x) \delta A_{\beta}(y)} + \left\langle {T^\mu}_\mu(z) V^{\alpha}(x) V'^{\beta}(y) 
\right\rangle.
\label{traceid1}
\eeq
Here $\mathcal A(z)$ is the anomaly functional, while $A_{\alpha}$ indicates the gauge fields coupled to the current $V^{\alpha}$.  $\Gamma^{\alpha\beta}$ is a generic 
dilaton/gauge/gauge vertex, which is obtained form the $TVV'$ vertex by tracing the spacetime indices $\mu\nu$.  $\mathcal A(z)$ is derived from the renormalized expression 
of the vertex by tracing the gravitational counterterms in $4-\epsilon$ dimensions (see for instance 
\cite{CDMS})
\beqa
\langle T_\mu^\mu \rangle=\mathcal A(z),
\eeqa
which in a curved background is given by the metric functional 
\beqa
\mathcal A(z) -\frac{1}{8} \left[ 2 b \,C^2 + 2 b' \left( E - \frac{2}{3}\square R\right) + 2 c\, F^2\right],
\label{anomalyeq}
\eeqa
 where $b$, $b'$ and $c$ are parameters.  For the case of a single fermion in an abelian gauge theory they are: $b = 1/320 \, \pi^2$,  $b' = - 11/5760 \, \pi^2$,
and $c= -e^2/24 \, \pi^2$. $C^2$ is the square of the Weyl tensor and $E$ is the Euler density given by
\beqa
C^2 &=& C_{\lambda\mu\nu\rho}C^{\lambda\mu\nu\rho} = R_{\lambda\mu\nu\rho}R^{\lambda\mu\nu\rho}
-2 R_{\mu\nu}R^{\mu\nu}  + \frac{R^2}{3}  \\
E &=& ^*\hskip-.1cm R_{\lambda\mu\nu\rho}\,^*\hskip-.1cm R^{\lambda\mu\nu\rho} =
R_{\lambda\mu\nu\rho}R^{\lambda\mu\nu\rho} - 4R_{\mu\nu}R^{\mu\nu}+ R^2.
\eeqa
In a flat metric background the expression of such functional reduces to the simple form
\beqa \label{TraceAnomaly}
\mathcal A(z)
&=& \sum_{i} \frac{\beta_i}{2 g_i} \, F^{\alpha\beta}_i(z) F^i_{\alpha\beta}(z), 
\eeqa
where $\beta_i$ are clearly the mass-independent $\beta$ functions of the gauge fields
and $g_i$ the corresponding coupling constants. For an extension which is quantum conformal invariant, the $\beta_i$ vanish. 

\begin{figure}[t]
\centering
\subfigure[]{\includegraphics[scale=.7]{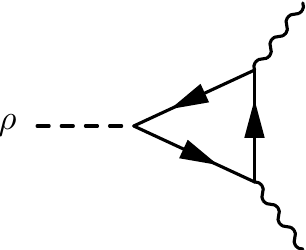}}
\hspace{.2cm}
\subfigure[]{\includegraphics[scale=.7]{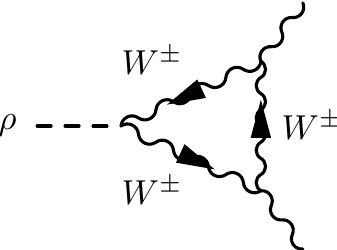}}
\hspace{.2cm}
\subfigure[]{\includegraphics[scale=.7]{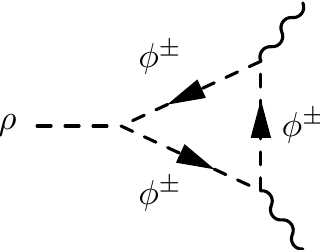}}
\hspace{.2cm}
\subfigure[]{\includegraphics[scale=.7]{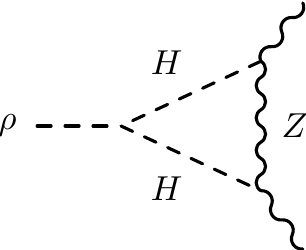}}
\centering
\subfigure[]{\includegraphics[scale=.7]{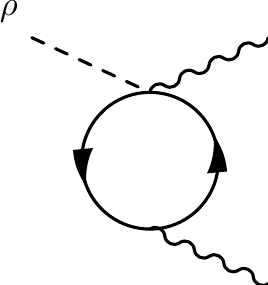}}
\hspace{.2cm}
\subfigure[]{\includegraphics[scale=.7]{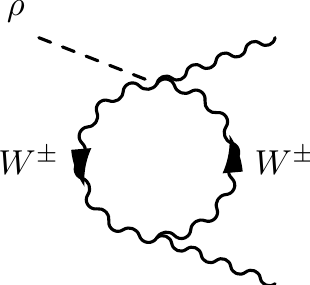}}
\hspace{.2cm}
\subfigure[]{\includegraphics[scale=.7]{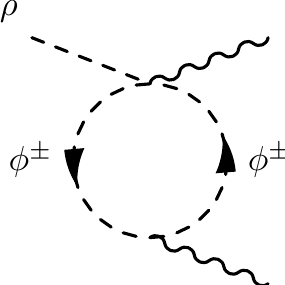}}
\hspace{.2cm}
\subfigure[]{\includegraphics[scale=.7]{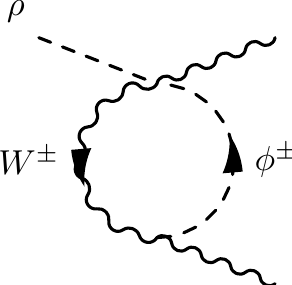}}
\subfigure[]{\includegraphics[scale=.7]{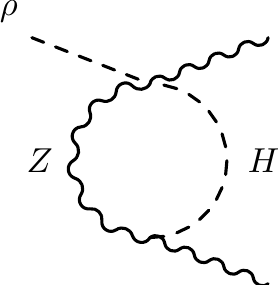}} 
\subfigure[]{\includegraphics[scale=.7]{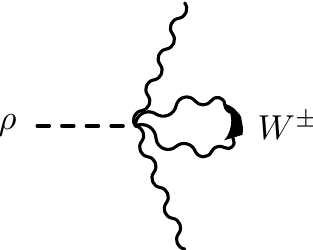}}
\caption{Typical amplitudes of triangle and bubble  topologies contributing to the $\rho \gamma\gamma$, $\rho \gamma Z$ and $\rho ZZ$ interactions. They 
include fermion $(F)$, gauge bosons $(B)$ and contributions from the term of improvement (I). Diagrams (a)-(g) contribute to all the 
three channels while (h)-(k) only in the $\rho ZZ$ case.}
\label{figuretriangle}
\end{figure}

\begin{figure}[t]
\centering
\subfigure[]{\includegraphics[scale=.7]{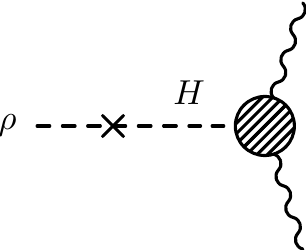}}
\hspace{.2cm}
\subfigure[]{\includegraphics[scale=.7]{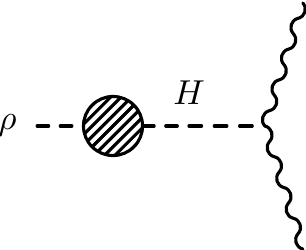}}
\hspace{.2cm}
\subfigure[]{\includegraphics[scale=.7]{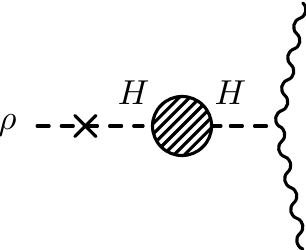}}
\caption{External leg corrections. Diagrams (b) and (c) appear only in the $\rho Z Z$ sector.}
\label{figuremix}
\end{figure}

The two terms on the right hand side of (\ref{traceid1}) are identified 
by computing the renormalized vertex $\langle T^{\mu\nu} V^\alpha V'^\beta\rangle$ and its trace.  It can be checked that the insertion of the (tree-level) trace of $T^{\mu\nu}$ into a two point function $VV'$, allows to 
identify the second term on the right-hand-side of the same equation, $\langle T^{\mu}_\mu(z) V^{\alpha}(x) V'^{\beta}(y)\rangle$. The difference between the trace of the lhs of 
(\ref{traceid1}) - which is computed from the correlator with open indices - and the vertex 
obtained by the direct insertion of  $T^\mu_\mu$, corresponds to the anomaly. It reproduces the $\mathcal A$-term, obtained by differentiating twice the  anomaly functional $\mathcal A$ with respect to the external source (the gauge field) \cite{CDMS}.\\
Beside the contribution from the anomaly,  the remaining contributions are contained, for each decay channel, into 2 additional form factors, denoted as  $\Sigma$ and $\Delta$. $\Sigma$ and 
$\Delta$ terms are related to the exchange of fermions, gauge bosons and scalars (Higgs/Goldstones). Explicit results for the $\rho VV'$ vertices ($V,V' = \gamma, Z$), denoted as $\Gamma_{VV'}^{\alpha \beta}$, are given in \cite{CDS}
which are decomposed in momentum space in the form 
\beq
\Gamma_{VV'}^{\alpha \beta}(k,p,q) = (2\, \pi)^4\, \delta^4(k-p-q) \frac{i}{\Lambda} 
\left( \mathcal A^{\alpha \beta}(p,q) + \Sigma^{\alpha \beta}(p,q) + \Delta^{\alpha \beta}(p,q)\right),
\eeq
where 
\beq
\mathcal A^{\alpha \beta}(p,q) = \int d^4 x\, d^4 y \, e^{i p \cdot x + i q\cdot y}\, 
\frac{\delta^2 \mathcal A(0)}{\delta A^\alpha(x)\delta A^\beta(y)}
\eeq
and 
\beq
 \Sigma^{\alpha \beta}(p,q) +  \Delta^{\alpha \beta}(p,q) = \int d^4 x\, d^4 y\, e^{ i p \cdot x + i q\cdot y}\, 
\left\langle {T^\mu}_\mu(0) V^\alpha(x) V^\beta(y) \right\rangle \,.
\eeq
Typical contributions are shown in Fig. \ref{figuretriangle}. We have denoted with $\Sigma^{\alpha \beta}$ the cut vertex contribution to $\Gamma^{\alpha\beta}_{\rho VV'}$, 
while $\Delta^{\alpha \beta}$ includes the dilaton-Higgs mixing on the dilaton line, as shown in Fig. \ref{figuremix}. The bilinear mixing $\Delta^{\a\b}$ does not appear in the decay amplitude, since this has to be cut on the external lines, but it plays a role in the overall renormalization of the effective theory. If the dilaton is described by a conformally coupled scalar, then the one-loop renormalization of the SM Lagrangian is sufficient for removing all the singularities present in this vertex, and specifically, in the bilinear mixing \cite{CDS}. For a dilaton described by a generic non-minimal/minimally coupled scalar, 
then this 2-point function contributions $\Delta$ requires an extra counterterm, generated by the renormalization of the term of improvement. A complete study of the $TVV'$ vertex and of the relative Ward and Slavnov-Taylor (STI) identities which can be used to secure the  correctness of the complete perturbative result can be found in \cite{Coriano:2011zk} for the electroweak theory. The analysis in QED and QCD can be found in \cite{Giannotti:2008cv, Armillis:2009pq} and \cite{ACD2}, respectively.

\subsection{The coupling to the anomaly and the breaking of quantum scale invariance}
As we have mentioned above, for a classical scale invariant estension, the coupling of the 
dilaton to the fields of the SM is characterised by two terms, the first of them being proportional to the anomaly. In the case of a quantum scale invariant extension 
\cite{GGS}, this term is obviously absent, due to a vanishing beta functions, but it reappears as an effective interaction if the fermions of the high energy spectrum of the quantum conformal theory are far heavier than the scale at which we probe the theory, which in this case is the LHC scale. This simple phenomenon can be easily understood in perturbation theory by looking at the fermion sector of the $\rho$/gauge/gauge vertex, for on shell external gauge lines. The corresponding triangle diagram is expressed from the standard one-loop scalar integral $C_0(s,m_i^2)$, where $s$ is of the order of the dilaton mass, and $m_i$ the mass of each particle running in the loop. The corresponding interaction takes the form 
\beq
\Gamma_{\rho V V}\sim \frac{g^2}{\pi^2 \Lambda} \, m_i^2 \, 
\bigg[ \frac{1}{s}  -  \frac{1} {2 } C_0 (s, m_i^2) \bigg(1-\frac{4 m_i^2}{ s}\bigg)\bigg] 
\sim 
\frac{g^2}{\pi^2 \Lambda} \, \frac{1}{6} + O\left( \frac{s}{m_i^2} \right),
\eeq
where  we have performed the large mass limit of the amplitude $(m_i \gg s)$ 
using 
\beqa
C_0(s, m_i^2) \sim - \frac{1}{2 m_i^2} \left( 1 + \frac{1}{12} \frac{s}{m_i^2} + O(\frac{s^2}{m_i^4} ) \right).
\label{dec}
\eeqa
 This shows that in the case of heavy fermions, 
the dependence on the fermion mass cancels, with the appearance of a point-like coupling of the dilaton to the trace anomaly $FF$.

Obviously, this limit generates an effective coupling which is proportional to the $\beta$ function related to the heavy flavours. On the other hand, the complete 
$\beta$ functions, including the contribution from all states, must vanish
\bea
\beta = \frac{g^3}{16 \pi^2} �\bigg[ \sum_{i} b^i �+ \sum_{j} b^j \bigg] = 0 \,,
\eea
where $i$ and $j$ run over the heavy and light states respectively. Exploiting the consequence of the quantum conformal symmetry, the 
dilaton couplings to the massless gauge bosons become
\bea
\mathcal L_{\rho} = 
- \frac{\alpha_s}{8 \pi} \sum_j b_g^j \, \frac{\rho}{\Lambda} (F_{g \, \mu\nu}^a)^2 � - � \frac{\alpha_{em}}{8 \pi}
\sum_j b_{em}^j \, \frac{\rho}{\Lambda} (F_{\gamma \, \mu\nu})^2 � \,,
\eea
in which the dependence on the $\beta$ functions of the light states is now explicit.  The appearance of the light 
states contributions to the $\beta$ functions is a consequence of the vanishing of the complete $\beta$ function and of the decoupling mechanism summarised by the loop behaviour in 
(\ref{dec}). \\

 \section{Decays of the dilaton}\label{decays}
\begin{figure}[t]
\begin{center}
\hspace*{-2cm}
\mbox{\subfigure[]{
\includegraphics[width=0.6\linewidth]{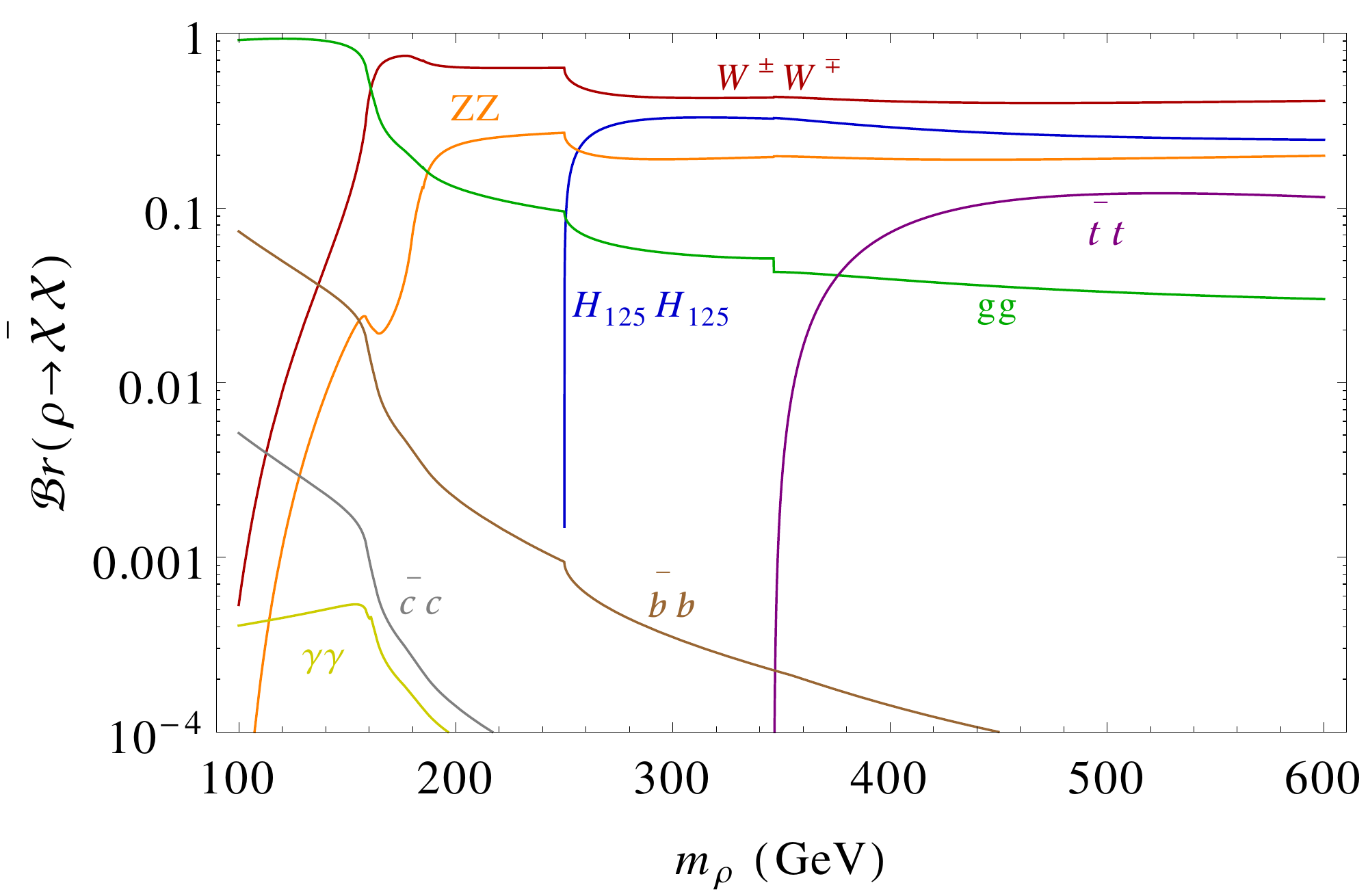}}\hskip 15pt
\subfigure[]{\includegraphics[width=0.6\linewidth]{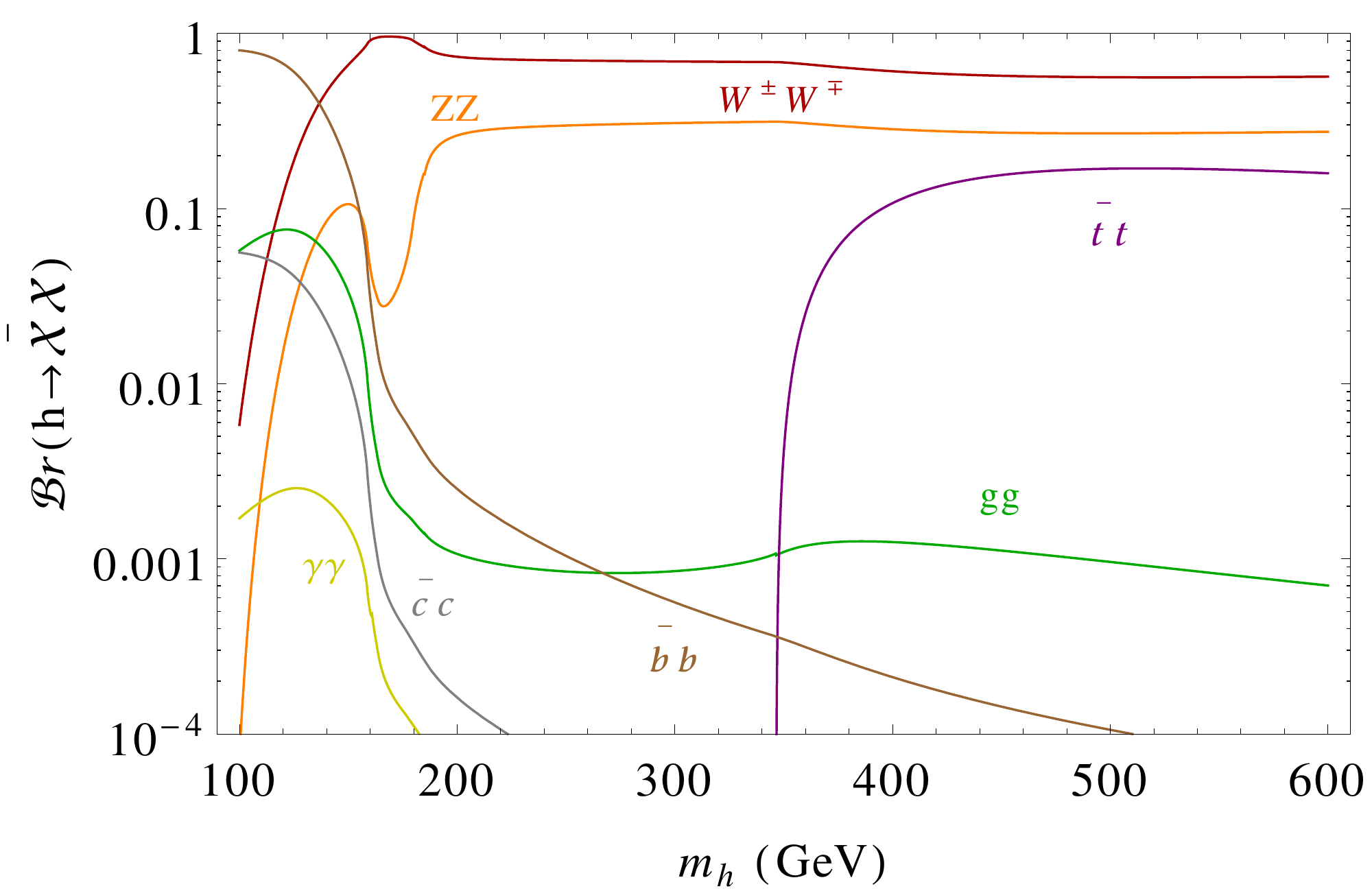}}}
\caption{The mass dependence of the branching ratios of the dilaton (a) and of the Higgs boson (b).}\label{brrhoh}
\end{center}
\end{figure}

 We start considering the case where there is no bilinear mixing between the Higgs and dilaton ($\xi=0$).  The interactions of the dilaton to the massive states are very similar to those of the Higgs, except that $v$ is replaced by $\Lambda$. The distinctive feature between the dilaton and the SM Higgs emerges in the coupling with photons and gluons.
One-loop expressions for the decays into all the neutral currents sector has been given in \cite{CDS}, while leading order decay widths of $\rho$ in some relevant channels (fermions, vector and Higgs pairs) are easily written in the form (for a minimally coupled dilaton, with $\xi=0$)
\begin{align}
&\Gamma_{\rho\to \bar ff}=N_f^c\frac{m_\rho}{8\pi}\frac{m_f^2}{\Lambda^2}\left(1-4\frac{m_f^2}{m_\rho^2}\right)^{3/2},\label{ffWidth}\\
&\Gamma_{\rho\to VV}=\delta_V\frac{1}{32\pi}\frac{m_\rho^3}{\Lambda^2}\left(1-4\frac{m_V^2}{m_\rho^2}+12\frac{m_V^4}{m_\rho^4}\right)\sqrt{1-4\frac{m_V^2}{m_\rho^2}},\label{vvWidth}\\
&\Gamma_{\rho\to HH}=\frac{1}{32\pi}\frac{m_\rho^3}{\Lambda^2}\left(1+2\frac{m_H^2}{m_\rho^2}\right)^2\sqrt{1-4\frac{m_H^2}{m_\rho^2}}\label{hhWidth}.
\end{align}
The one-loop expression for decays into $\gamma\gamma$ is
\beqa
\Gamma(\rho \rightarrow \gamma\gamma) 
&=&
\frac{\alpha^2\,m_{\rho}^3}{256\,\Lambda^2\,\pi^3} \, \bigg| \beta_{2} + \beta_{Y} 
 -\left[ 2 + 3\, x_W  +3\,x_W\,(2-x_W)\,f(x_W) \right] \nn \\
&& + \frac{8}{3} \, x_t\left[1 + (1-x_t)\,f(x_t) \right] \bigg|^2. \,
\label{PhiGammaGamma}\nn\\ 
\eeqa
Here, the contributions to the decay, beside the anomaly term, come from the $W$ and the fermion (top) loops. $\beta_2 (= 19/6)$ and $\beta_Y (= -41/6)$ are the $SU(2)_L$ and $U(1)_Y$ $\beta$ functions, while the $x_i$'s  are proportional to the ratios between the mass of each  particle in the loops $m_i$ and the $\rho$ mass. In general, we have defined the variable 
\beq \label{x}
x_i = \frac{4\, m_i^2}{m^2_\rho} \, ,
\eeq
with the index "$i$" labelling the corresponding massive virtual particles. The leading fermionic contribution in the loop comes from the top quark via $f(x_t)$, while $f(x_W)$ denotes the contribution of the $W$-loop. The function $f(x)$ is given by
\beqa
\label{fx}
f(x) = 
\begin{cases}
\arcsin^2(\frac{1}{\sqrt{x}})\, , \quad \mbox{if} \quad \,  x \geq 1 \\ 
-\frac{1}{4}\,\left[ \ln\frac{1+\sqrt{1-x}}{1-\sqrt{1-x}} - i\,\pi \right]^2\, , \quad \mbox{if} \quad \, x < 1.
\end{cases}
\eeqa
related to the scalar three-point master integral through the relation 
\beq \label{C03m}
C_0(s,m^2) = - \frac{2}{s} \, f(\frac{4\,m^2}{s}) \, .
\eeq
The decay rate of a dilaton into two gluons is given by
\beqa
\Gamma(\rho \rightarrow gg) 
&=&
\frac{\alpha_s^2\,m_\rho^3}{32\,\pi^3 \Lambda^2} \, \bigg| \beta_{QCD} +  x_t\left[1 + (1-x_t)\,f(x_t) \right] \bigg|^2 \,,
\label{ggWidth}
\eeqa
where $\beta_{QCD}$ is the QCD $\beta$ function and we have taken the top quark as the only massive fermion, with $x_i$ and $f(x_i)$ defined in Eq. (\ref{x}) and Eq. (\ref{fx}) respectively. \\
Differently from the cross section case,
 the dependence of the decay amplitudes Eq.~(\ref{ffWidth}) - Eq.~(\ref{hhWidth}) on the conformal scale $\Lambda$, which amounts to an overall factor,  
the branching ratios
\bea
\mathit{Br}(\rho\to\bar X X)=\frac{\Gamma_{\rho\to\bar X X}}{\sum_X\Gamma_{\rho\to \bar X X}},
\eea
are $\Lambda$-independent. \\
We show in Fig.~\ref{brrhoh}(a) the decay branching ratios of the dilation as a function of its mass, while in Fig.~\ref{brrhoh}(b) we plot the  corresponding decay branching ratios for a SM-like heavy Higgs boson, here assumed to be of a variable mass. For a light dilaton with $m_\rho < 200$ GeV the dominant decay mode is into two gluons ($gg$), while for a dilaton of larger mass ($m_\rho > 200$ GeV) the same channels which are available for the SM-like Higgs ($ZZ, WW, \bar{t} t$) are now accompanied by a significant $gg$ mode. From the two figures it is easily observed that the 2 gluon rate in the Higgs case is at the level of few per mille, while in the dilaton case is just slightly below 10$\%$.

\section{Production of the dilaton}\label{prod}
The main production process of the dilaton at the LHC is through gluon fusion, as for the Higgs boson, with a suppression induced by the conformal breaking scale $\Lambda$, which lowers the production rates. Even in this less favourable situation, if confronted with the Higgs production rates of the SM, the dilaton phenomenology can still be studied al the LHC. \\
We calculate the dilaton production cross-section via gluon fusion by weighting the Higgs boson to gluon-gluon decay widths with the corresponding dilaton decay width. The dilaton production cross-section with the incoming gluons thus can be written as
\bea\label{ggrho}
\sigma_{gg\to\rho}=\sigma_{gg\to H}\,\,\frac{\Gamma_{\rho\to gg}}{\Gamma_{H\to gg}} ,
\eea
where we use the same factorization scale in the DGLAP evolution of the parton distribution functions (PDF) of \cite{HCWG}.
The width of $\rho\to gg$ is given in Eq.~(\ref{ggWidth}) and we can use the same expression to calculate the width of $H\to gg$, replacing the breaking scale $\Lambda$ with $v$ and setting $\beta_{QCD}\equiv 0$. The ratio of the two widths appearing in Eq.~(\ref{ggrho}) is then given by
\bea\label{Wrhoh}
\frac{\Gamma_{\rho\to gg}}{\Gamma_{H\to gg}}=\frac{v^2}{\Lambda^2}\frac{m^3_\rho}{m^3_H}\frac{\left|\beta_{QCD}+ x_t\left[1 + (1-x_t)\,f(x_t)\right] \right|^2}{\left| x_t\left[1 + (1-x_t)\,f(x_t)\right] \right|^2} .
\eea

\begin{figure}[thb]
\begin{center}
\hspace*{-2cm}
\mbox{\subfigure[]{
\includegraphics[width=0.6\linewidth]{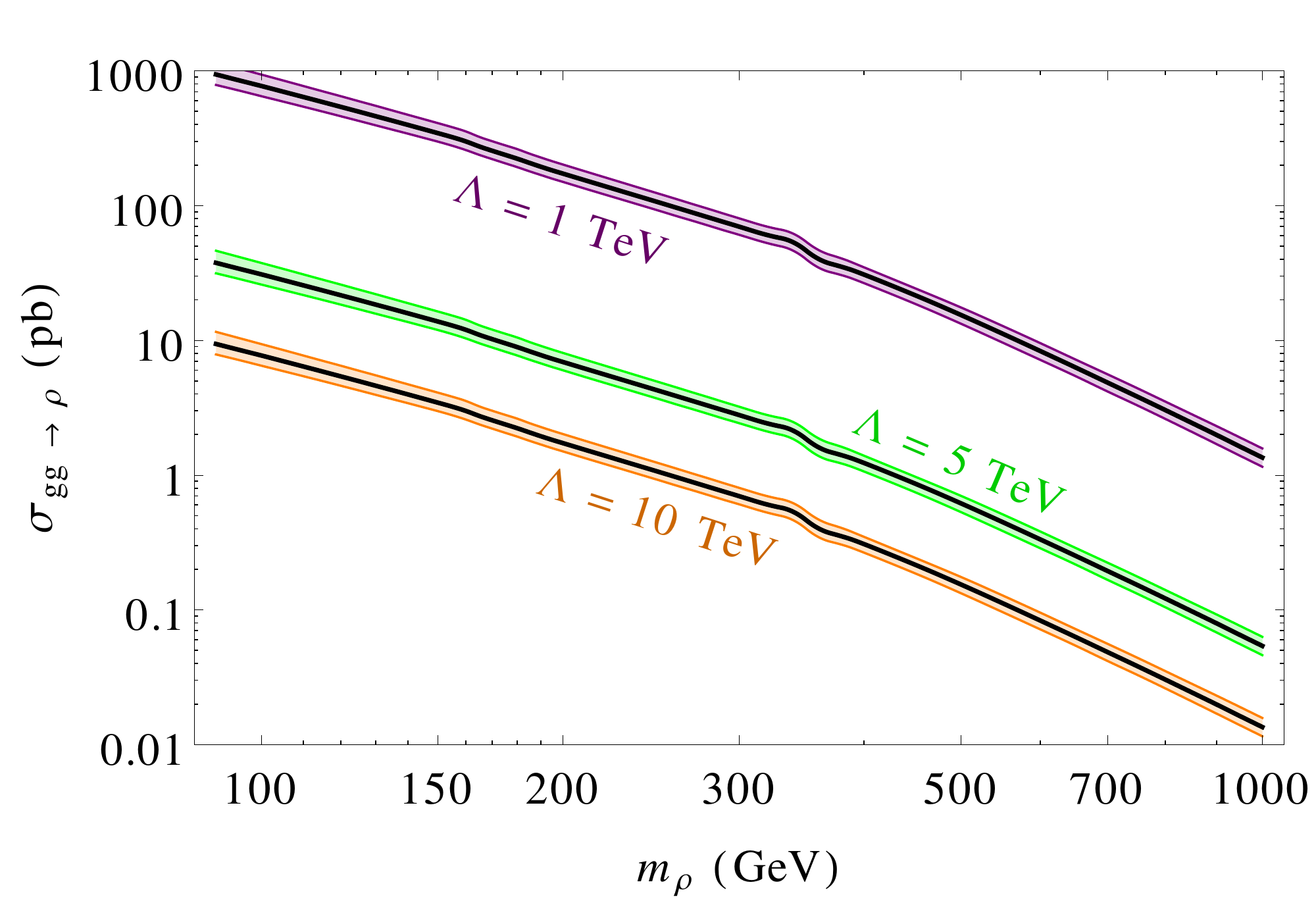}}\hskip 15pt
\subfigure[]{\includegraphics[width=0.6\linewidth]{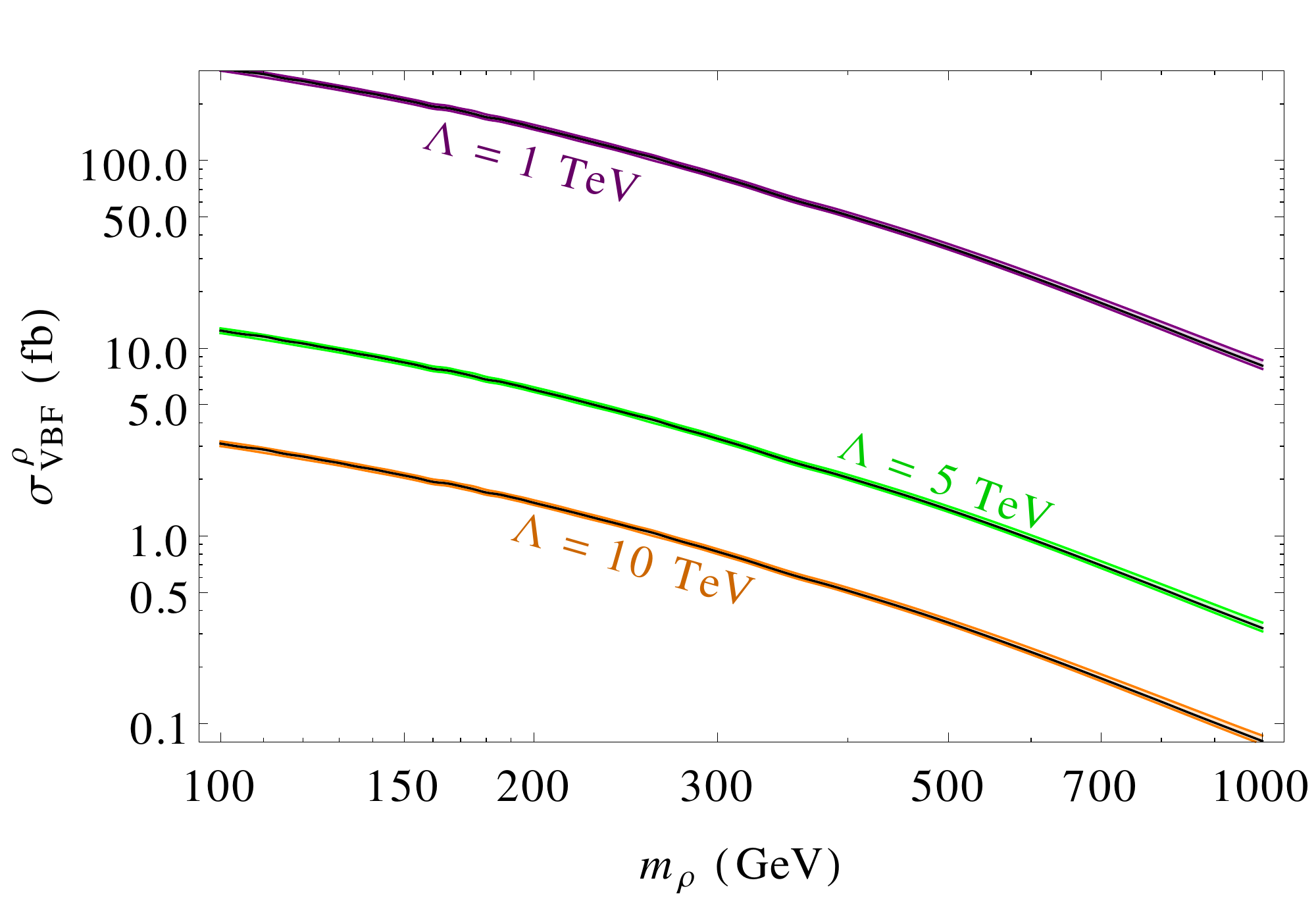}}}
\caption{The mass dependence of the dilaton cross-section via gluon fusion (a) and vector boson fusion (b) for three different choices of the conformal scale, $\Lambda=1, 5, 10$ TeV respectively.}\label{ggvbf}
\end{center}
\end{figure}

In Fig.~\ref{ggvbf} we present the production cross-section of the dilaton at the LHC at 14 TeV centre of mass energy mediated by (a) gluon fusion and (b) vector boson fusion, versus $m_\rho$. Shown are the variations of the same observables for three conformal breaking scales with $\Lambda=1, 5, 10$ TeV. Notice that the contribution from the gluon fusion is about a factor $10^4$ larger than the vector boson fusion.
\subsection{Bounds on the dilaton from heavy Higgs searches at the LHC}
\begin{figure}
\begin{center}
\hspace*{-2cm}
\mbox{\subfigure[]{
\includegraphics[width=0.6\linewidth]{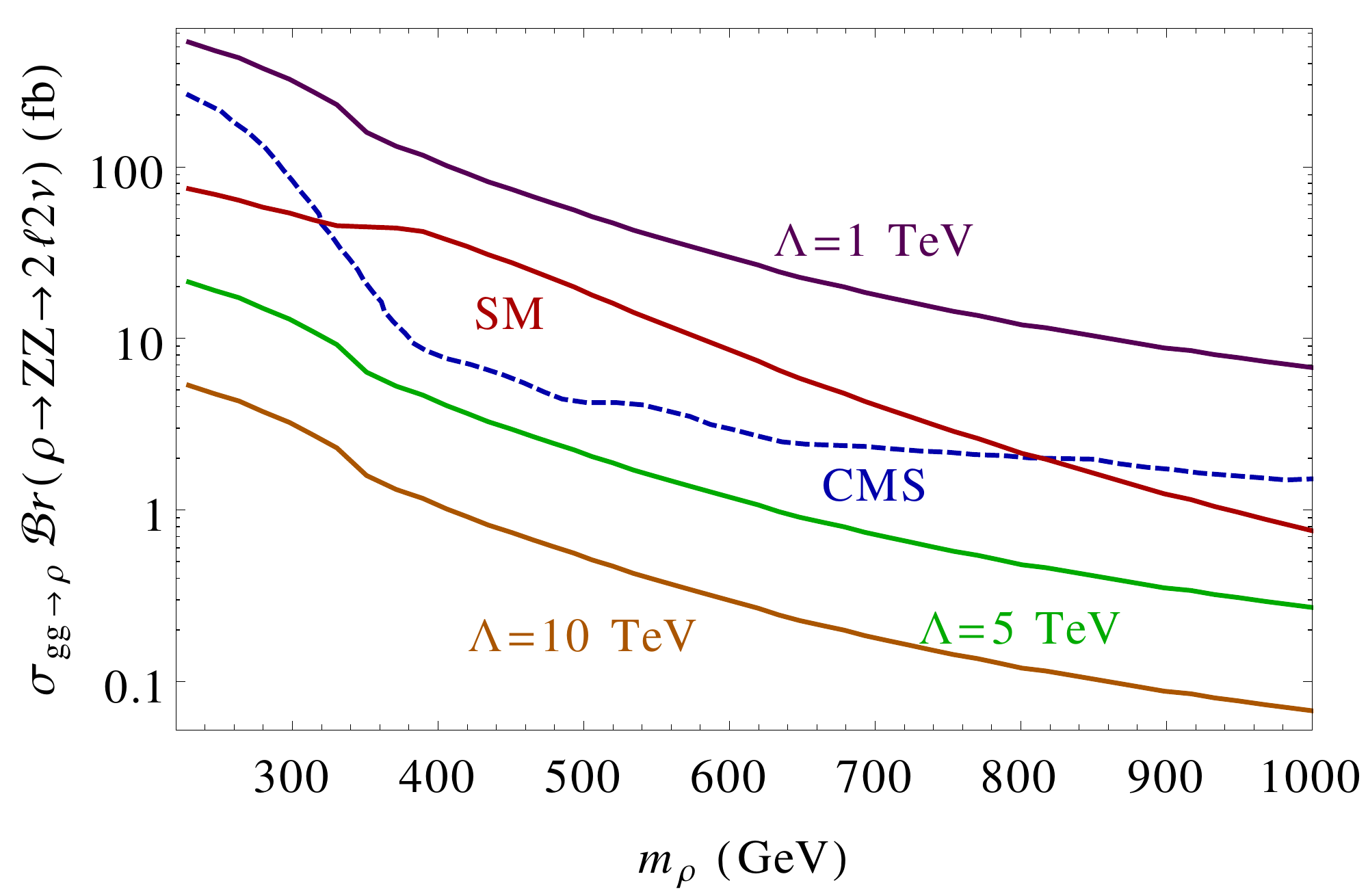}}\hskip 15pt
\subfigure[]{\includegraphics[width=0.6\linewidth]{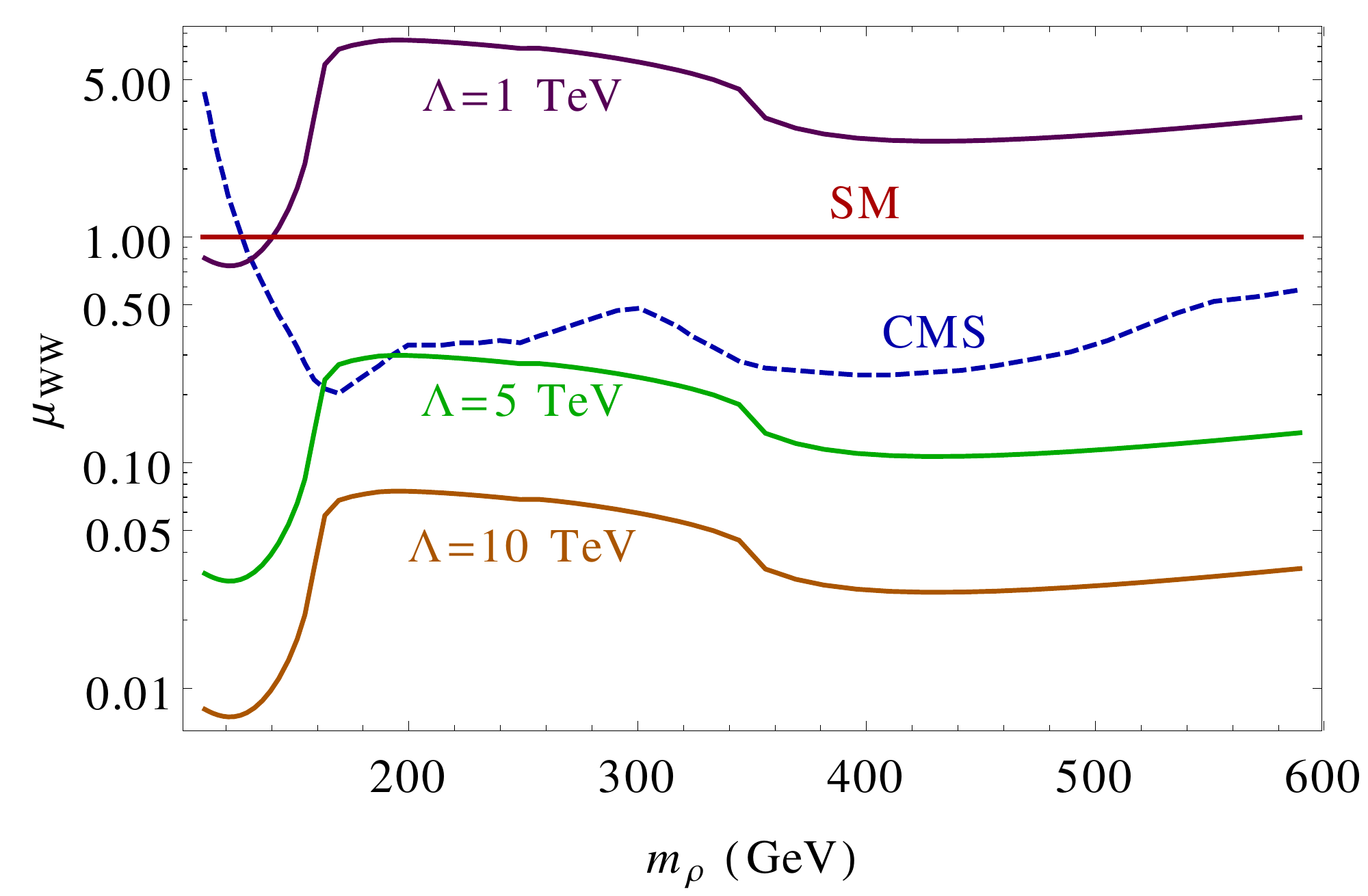}}}
\hspace*{-2cm}
\mbox{\subfigure[]{\includegraphics[width=0.6\linewidth]{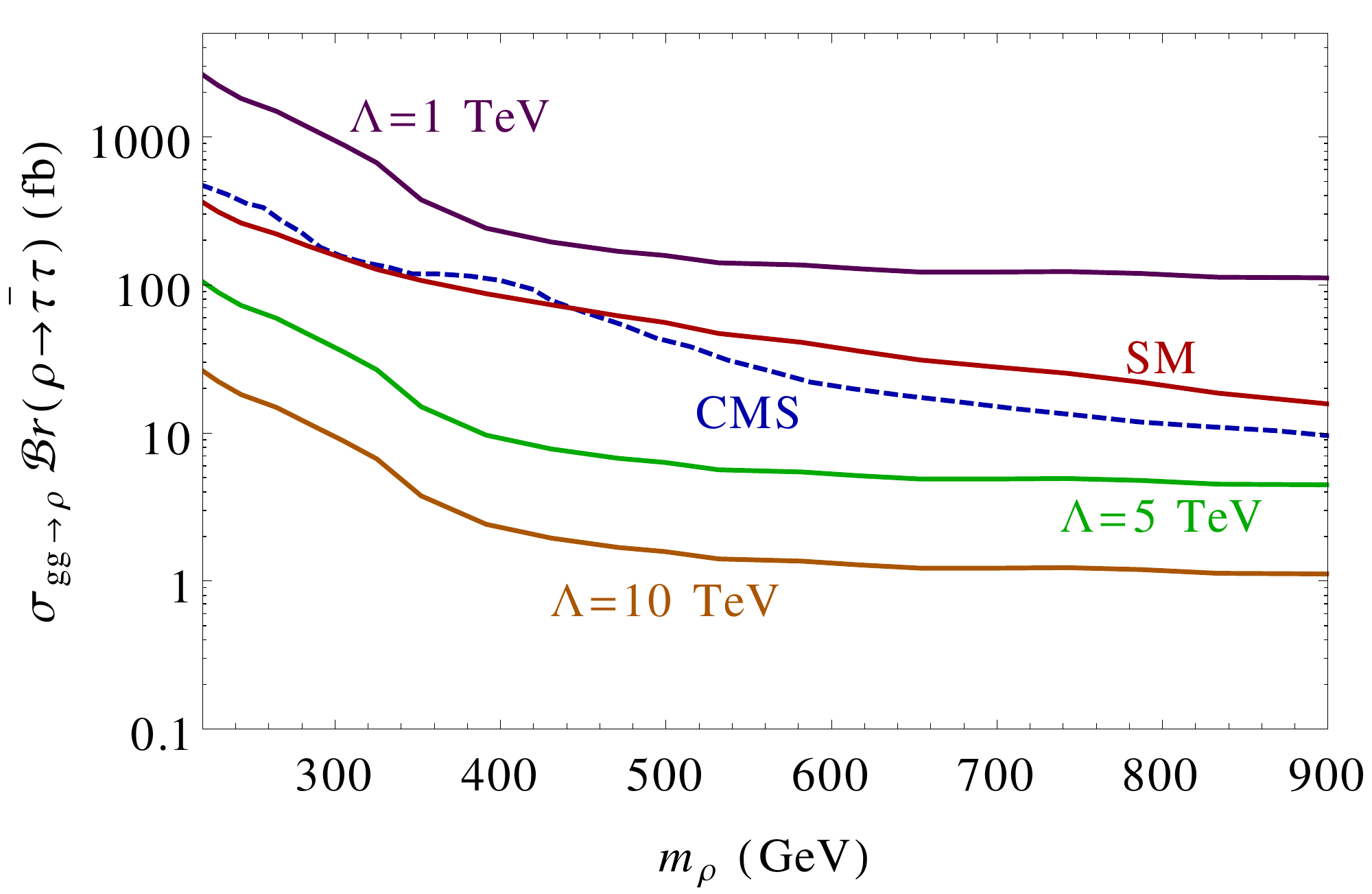}}\hskip 15pt
\subfigure[]{\includegraphics[width=0.6\linewidth]{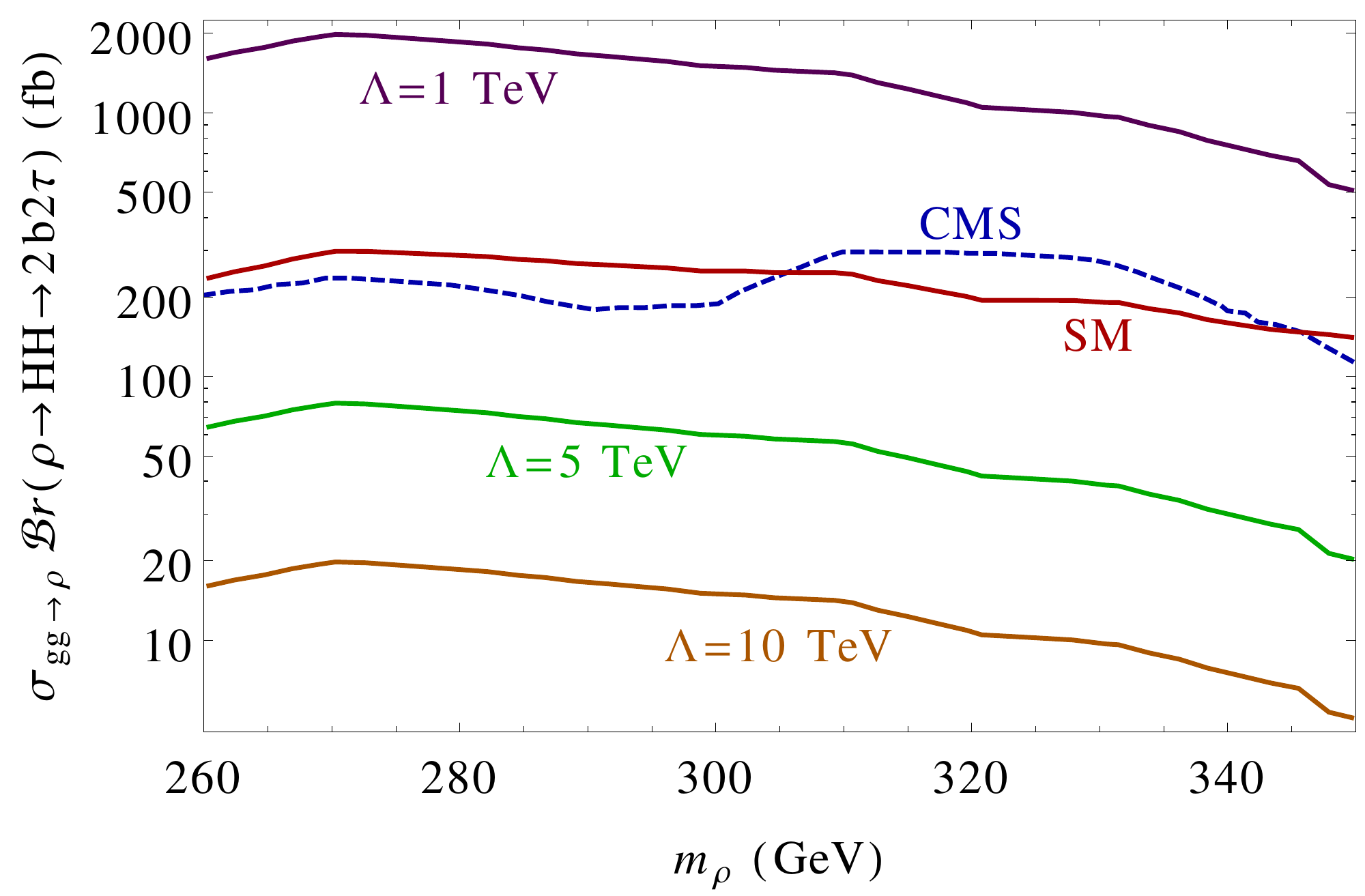}}}
\caption{The mass bounds on the dilaton from heavy scalar decays to (a) $ZZ$ \cite{CMSzz}, (b) $W^\pm W^\mp$ \cite{CMSww}, (c) $\bar\tau\tau$ \cite{CMStautau} and (d) to $H\,H$ \cite {CMShh} for three different choices of conformal scale, $\Lambda=1, 5, 10$ TeV respectively.}\label{bzzww}
\end{center}
\end{figure}
Since the mass of the dilaton is a free parameter, and given the similarities with the main  production and decay channels of this particle with the Higgs boson, several features of the production and decay channels in the Higgs sector, with the due modifications, are shared also by the dilaton case.  

As we have already mentioned, the production cross-section depends sensitively on $\Lambda$, as shown in Eqs.~(\ref{ggrho}) and (\ref{Wrhoh}). Bounds on this breaking scale has been imposed by the experimental searches for a heavy, SM-like Higgs boson at the LHC, heavier than the $125$ GeV Higgs, $H_{125}$.  \\
We have investigated the bounds on $\Lambda$ coming from the following datasets
\begin{itemize}
\item{}
the 4.9 $\rm{fb}^{-1}$ (at 7 TeV) and 19.7 $\rm{fb}^{-1}$ (at 8 TeV) datasets  for a heavy Higgs decaying into $Z\,Z$ \cite{CMSzz}, $W^\pm W^\mp$ \cite{CMSww}, $\bar\tau\tau$ \cite{CMStautau} and 
\item{}
the 19.7 fb$^{-1}$ datasets (at 8 TeV) for the decay in $H\,H$ \cite{CMShh} from CMS  
\item{}
the 20.3 fb$^{-1}$ at 8 TeV data from ATLAS for the decay of the heavy Higgs into $Z\,Z$ \cite{ATLASzz}
and $W^\pm W^\mp$ \cite{ATLASww}. 
\end{itemize}

\begin{figure}
\begin{center}
\hspace*{-2cm}
\mbox{\subfigure[]{
\includegraphics[width=0.6\linewidth]{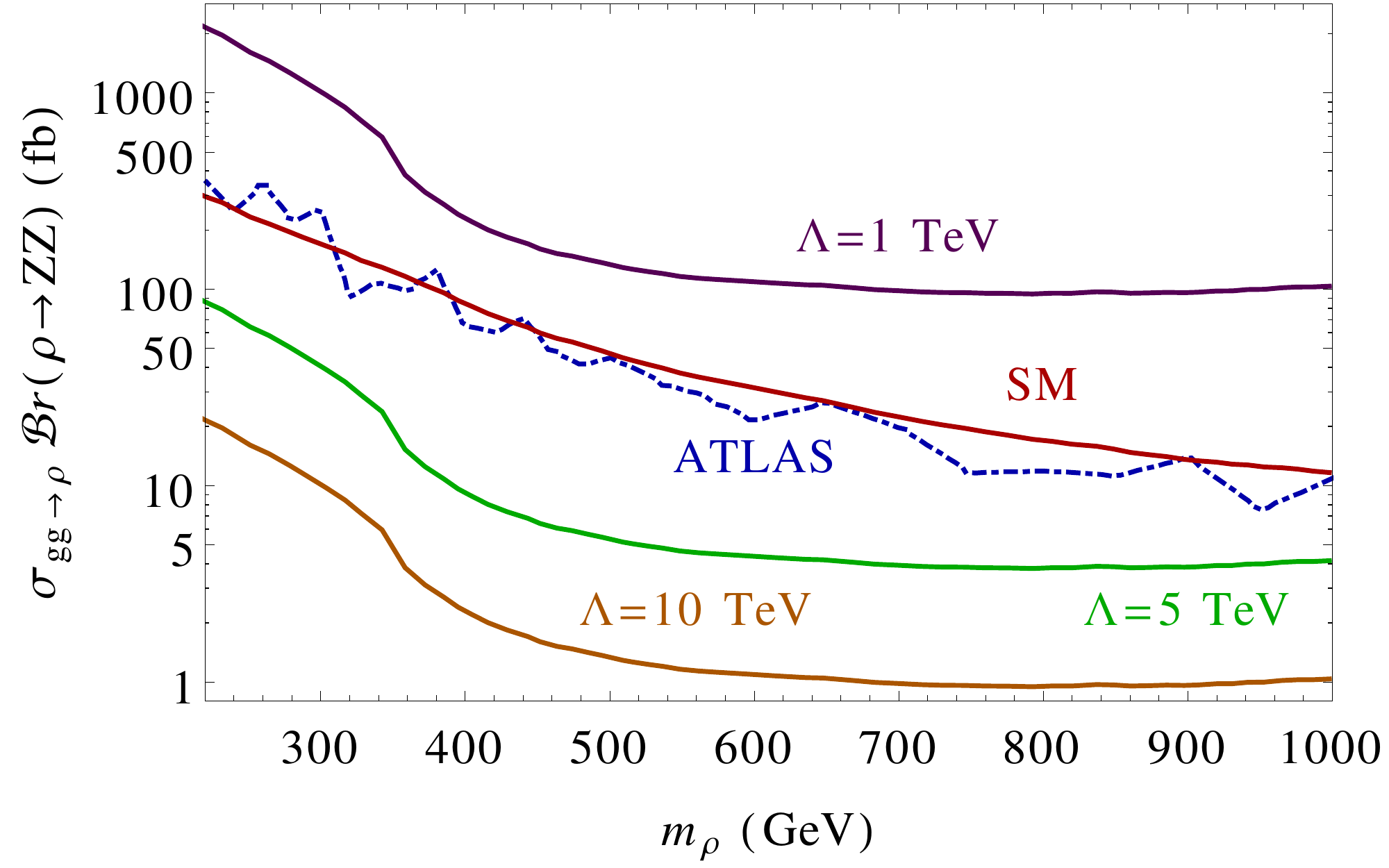}}\hskip 15pt
\subfigure[]{\includegraphics[width=0.6\linewidth]{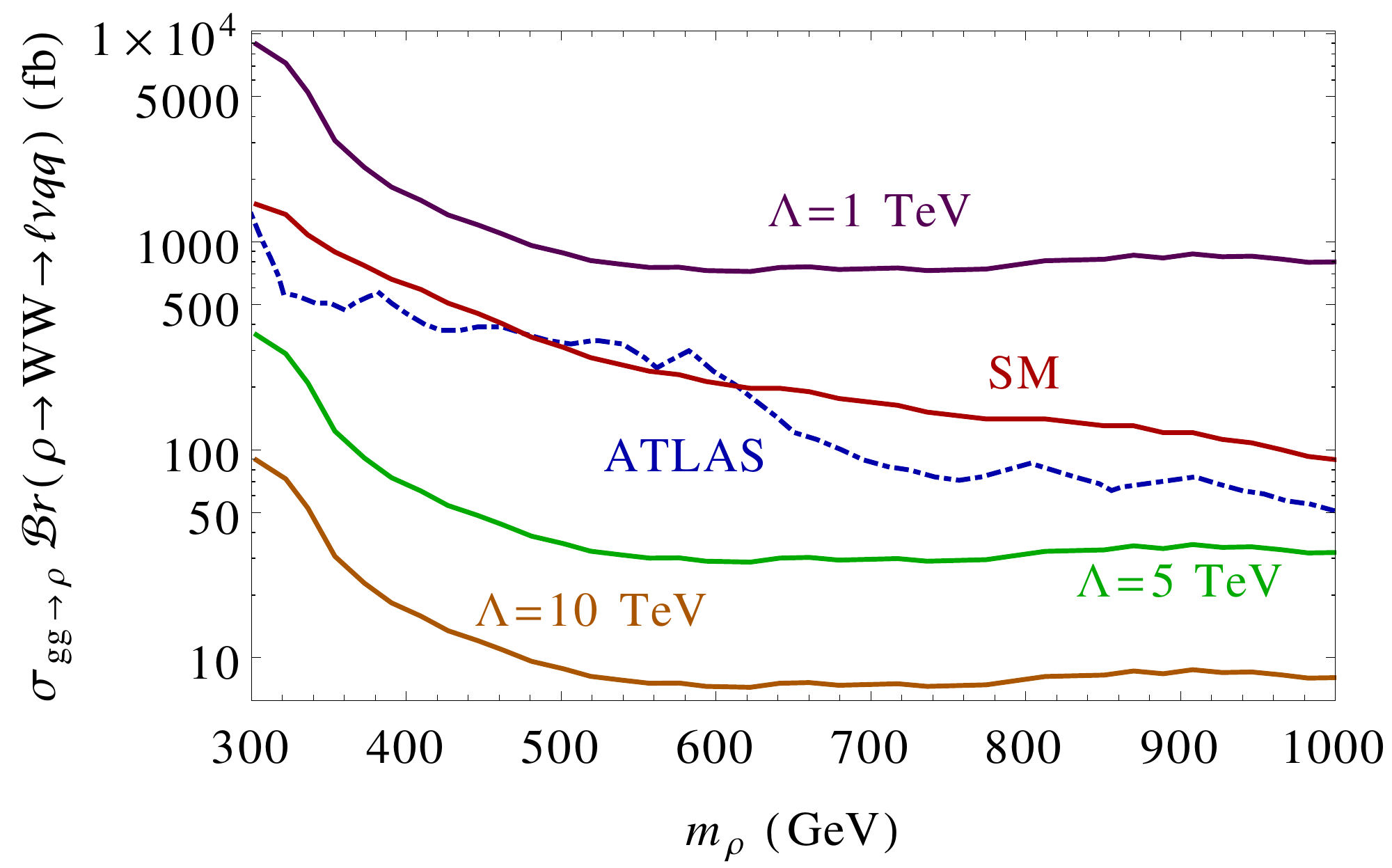}}}
\caption{The mass bounds on the dilaton from heavy scalar decays to (a) $ZZ$ \cite{ATLASzz} and (b) $W^\pm W^\mp$ \cite{ATLASww} for three different choices of conformal scale, $\Lambda=1, 5$ and $10$ TeV respectively.}\label{bzzwwA}
\end{center}
\end{figure}

The dotted line in each plot presents the upper bound on the cross-section, i.e. the $\mu$ parameter in each given modes defined as
\bea\label{mu}
\mu_{XY}=\frac{\sigma_{gg\to H}{{Br}(H\to XY)}}{{{\sigma_{gg \to H}}_{SM}}{{Br}(H \to XY)_{SM}}}.
\eea
In Fig. \ref{bzzww} we show the dependence of the 4-lepton ($2 l\, 2\nu$) channel on the mass of the $\rho$ at its peak, assuming $Z\,Z$, $W^\pm W^\mp$, $\bar\tau\tau$ and $H\,H$ intermediate states. 
The three continuous lines in violet, green and brown correspond to 3 diffferent values of the conformal scale, equal to 1, 5 and 10 TeV respectively. The SM predictions are shown in red. The dashed blue line 
separates the excluded and the admissible regions, above and below the blue curve respectively, which sets an upper bound of exclusion obtained from a CMS analysis.
A similar study is shown in  Fig. \ref{bzzwwA}, limited to the $Z\,Z$ and $W^\pm W^\mp$ channels, where we report the corresponding bound presented, in this case, by the ATLAS collaboration. Both the ATLAS and CMS data completely exclude the $\Lambda=1$ TeV case whereas the $\Lambda = 5$ TeV case has only a small tension with the CMS analysis of the $W^\pm W^\mp$ channel if $m_\rho\sim 160$ GeV.  Any value of $\Lambda \geq 5$ TeV is not  ruled out by the current data.\\
In Table \ref{cross} we report the values of the gluon fusion cross-section for three benchmark points 
(BP) that we have used in our phenomenological analysis. We have chosen $\Lambda = 5$ TeV, and the factorization in the evolution of the parton densities has been performed in concordance with those of the Higgs working group \cite{HCWG}. In the following subsection we briefly 
discuss some specific features of the dilaton phenomenology at the LHC, which will be confronted with a PYTHIA based simulation of the SM background.

 \begin{table}[t]
\begin{center}
\hspace*{-1.0cm}
\renewcommand{\arraystretch}{1.2}
\begin{tabular}{|c||c|c|c||}
\hline\hline
Benchmark& $m_\rho$& $gg\to \rho$\\
Points &GeV& in fb \\
\hline
BP1 &200&6906.62\\
\hline
BP2 &260&3847.45\\
\hline
BP3 &400&1229.25\\
\hline
\hline
\end{tabular}
\caption{Dilation production cross-section via gluon fusion at the LHC at 14 TeV, for the 3 selected benchmark points, with $\Lambda=5$ TeV.}\label{cross}
\end{center}
\end{table}

\subsection{Dilaton phenomenology at the LHC}
\begin{figure}[thb]
\begin{center}
\mbox{\subfigure[]{
\includegraphics[width=0.4\linewidth]{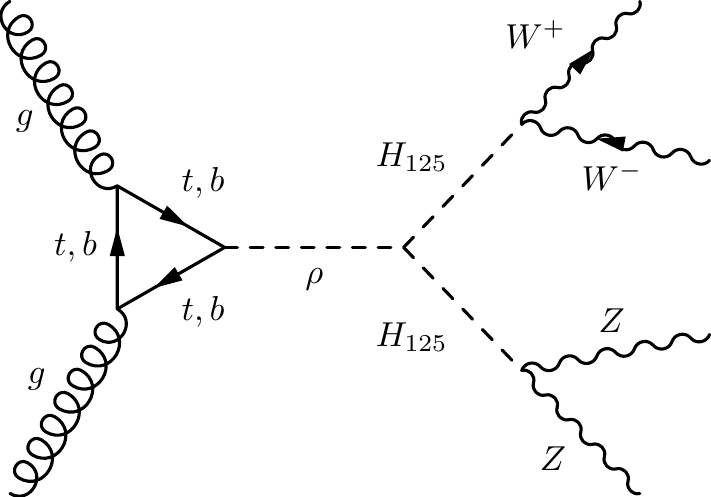}}
\hspace*{1cm}
\subfigure[]{\includegraphics[width=0.35\linewidth]{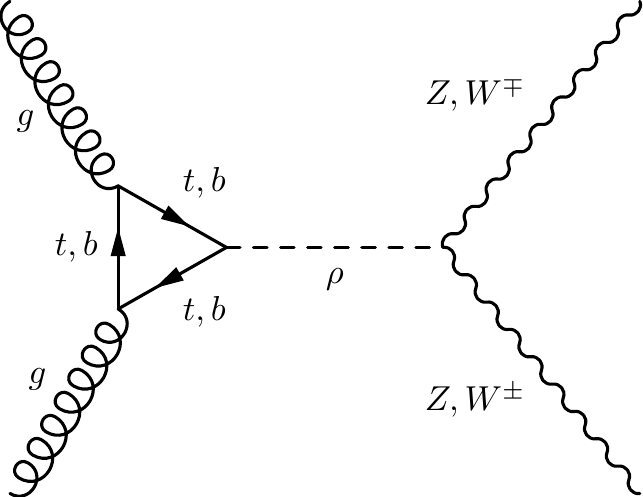}}}
\caption{The Feynman diagrams showing the dilaton production via gluon-gluon fusion and its decay to (a) pair of Higgs boson which further decays into gauge boson pairs and (b) a pair of gauge bosons.}\label{ggd}
\end{center}
\end{figure}

Fig.~\ref{ggd} shows the production and decay amplitudes mediated by an intermediate dilaton at the LHC.
We can see from Fig.~\ref{brrhoh}(a) that some of the main interesting decays of the dilaton are into two on-shell SM Higgs bosons $H\,H$, or into a real/virtual pair $H\,H^*$ and gauge boson pairs. The corresponding SM Higgs boson then further decays into $WW^*$ and/or $ZZ^*$.  Certainly these gauge bosons and their leptonic decays will give rise to multi-leptonic final states with missing transverse energy ($\etmiss$) via the chain 
\bea \label{dcy}
pp &\to& \rho \to  H\,H^* \nn\\
  &\to & WW^*, WW^* \nn \\
  & \to & 4\ell +\etmiss, \, 3\ell + 2j +\etmiss.
\eea
%
As shown above, there are distinct intermediate states mediating the decay of the dilaton into four $W^\pm$ bosons on/off-shell which give rise to $3\ell +\etmiss$ and $4\ell +2j +\etmiss$ final states. When we demand that one of the SM Higgs bosons $h$ decays to $ZZ^*$ and the other to $WW^*$, we gain a factor of two in multiplicity and generate a final state of the form $ 6\ell +\etmiss$, $ 4\ell + \geq 2j +\etmiss$ and $3\ell + 4j +\etmiss$ (i.e. 4 leptons, plus at least 2 jets accompanied by missing $E_T$) as in 
  \bea \label{dcy1}
pp &\to& \rho \to  H H^* \nn\\
  &\to & WW^*, ZZ^* \nn \\
  & \to & 6\ell +\etmiss, \, 4\ell + \geq 2j +\etmiss,\, 3\ell + 4j +\etmiss. 
\eea
Though the SM Higgs boson decay branching ratios to
$ZZ^*$ are relatively small $\sim 3\%$, when the dilaton decays via an intermediate $ZZ^*$, final states 
with several leptons are expected as in
\bea \label{dcy2}
pp &\to& \rho \to  H H^* \nn\\
  &\to & ZZ^*, ZZ^* \nn \\
  & \to & 8\ell,\, 6\ell + 2j,\, 4\ell + 4j. 
\eea
From the last decay channel, final states with multiple charged leptons and zero missing energy are now allowed, a case which we will explore next. \\
 The SM gauge boson branching ratios to charged leptons are very small, specially for channels mediated by a $Z$, due to the small rates. Therefore leptonic final states of higher multiplicities will be suppressed compared to those of a low number.  For this reason we will restrict the choice of the leptonic final states in our  simulation to $\geq 3\ell +X$ and $\geq 4\ell +X$. The requirement of 
$\geq 3\ell$ and $\geq 4\ell$  already allow to reduce most of the SM backgrounds,  although not completely, due to some some irreducible components, as we are going to discuss next.

 \section{Collider simulation}\label{colsim}
 We analyse dilaton production by gluon-gluon fusion, followed by its decay either to a pair of SM-like Higgs bosons ($\rho \to H_{125} H_{125}$) or to 
 a pair of gauge bosons ($WW$, $ZZ$). The $H_{125}$ thus produced will further decay into gauge boson pairs, i.e. $W^\pm W^\mp$ and $ZZ$, giving rise to mostly leptonic final states, as discussed above.  When the intermediate decays into one or more gauge bosons in the hadronic modes are considered, then we get leptons associated with extra jets in the final states. For $m_{\rho} < 2m_{H_{125}}$ the dilaton decays to two on-shell $H_{125}$ states are not kinematically allowed. In that case we consider its direct decay into gauge boson pairs, $W^\pm W^\mp, ZZ$. In the following subsections we consider the two case separately, where we analyze final states at the LHC at 14 TeV and simulate the contributions coming from the SM backgrounds. \\ 
For this goal we have implemented the model in SARAH \cite{sarah}, generated the model files for CalcHEP \cite{calchep}, later used to produce the decay file SLHA containing the decay rates and the corresponding mass spectra. The generated events have then been simulated with {\tt PYTHIA} \cite{pythia} via the the SLHA interface \cite{slha}. The simulation at hadronic level has been performed using the {\tt Fastjet-3.0.3} \cite{fastjet} with the {\tt CAMBRIDGE AACHEN} algorithm with a jet size $R=0.5$ for the jet formation, chosen according to the following criteria:
\begin{itemize}
  \item the calorimeter coverage is $\rm |\eta| < 4.5$

  \item minimum transverse momenta of the jets $ p_{T,min}^{jet} = 20$ GeV and the jets are ordered in $p_{T}$
  \item leptons ($\rm \ell=e,~\mu$) are selected with
        $p_T \ge 20$ GeV and $\rm |\eta| \le 2.5$
  \item no jet should be accompanied by a hard lepton in the event
   \item $\Delta R_{lj}\geq 0.4$ and $\Delta R_{ll}\geq 0.2$
  \item Since an efficient identification of the leptons is crucial for our study, we additionally require  
a hadronic activity within a cone of $\Delta R = 0.3$ between two isolated leptons. This is defined by the condition on the transverse momentum $\leq 0.15\, p^{\ell}_T$ GeV in the specified cone.

\end{itemize}

\subsection{Benchmark points}
\begin{figure}[bht]
\begin{center}
\hspace*{-2cm}
\mbox{\subfigure[]{
\includegraphics[width=0.6\linewidth]{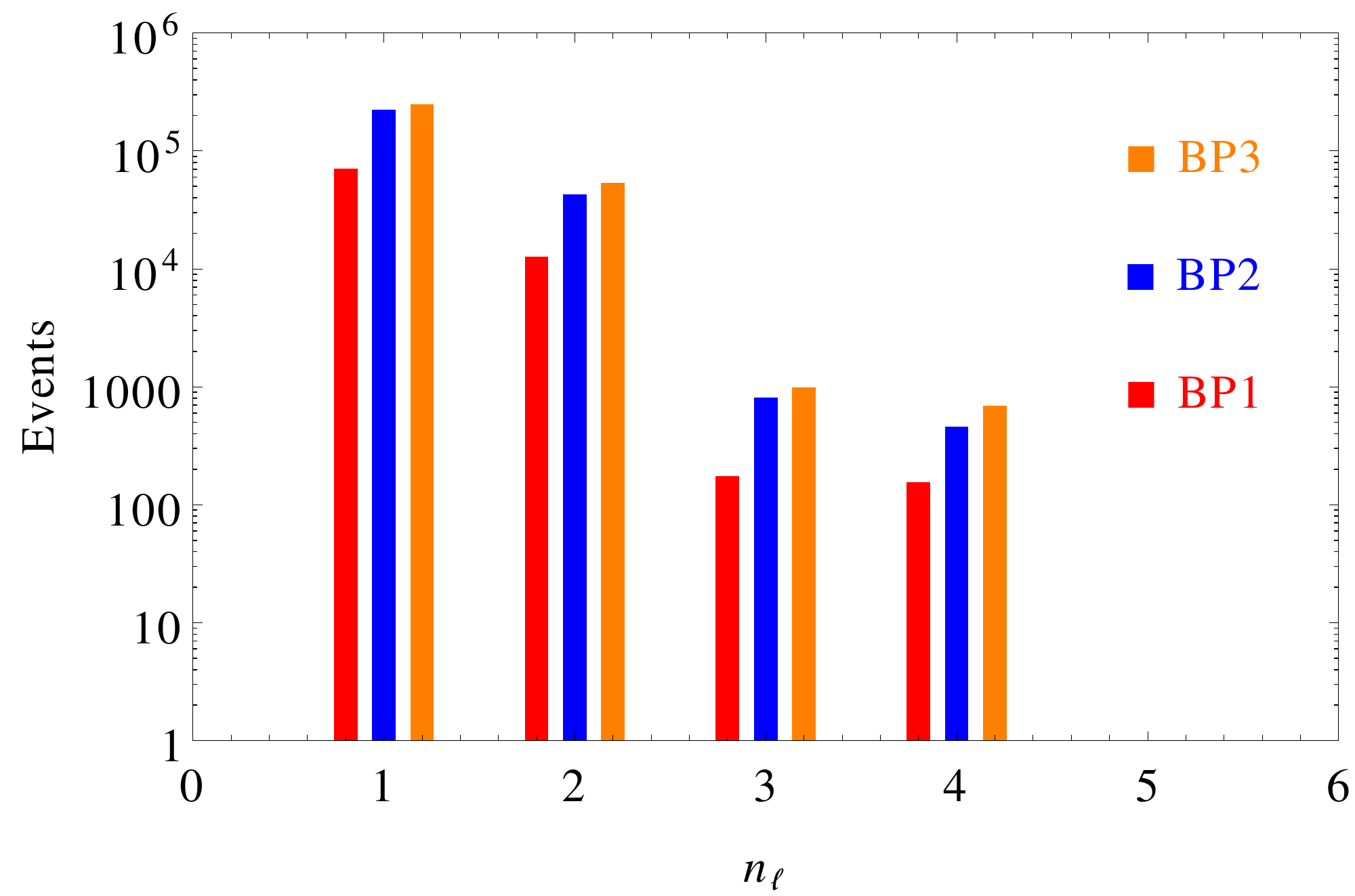}}
\hspace*{.5cm}
\subfigure[]{\includegraphics[width=0.6\linewidth]{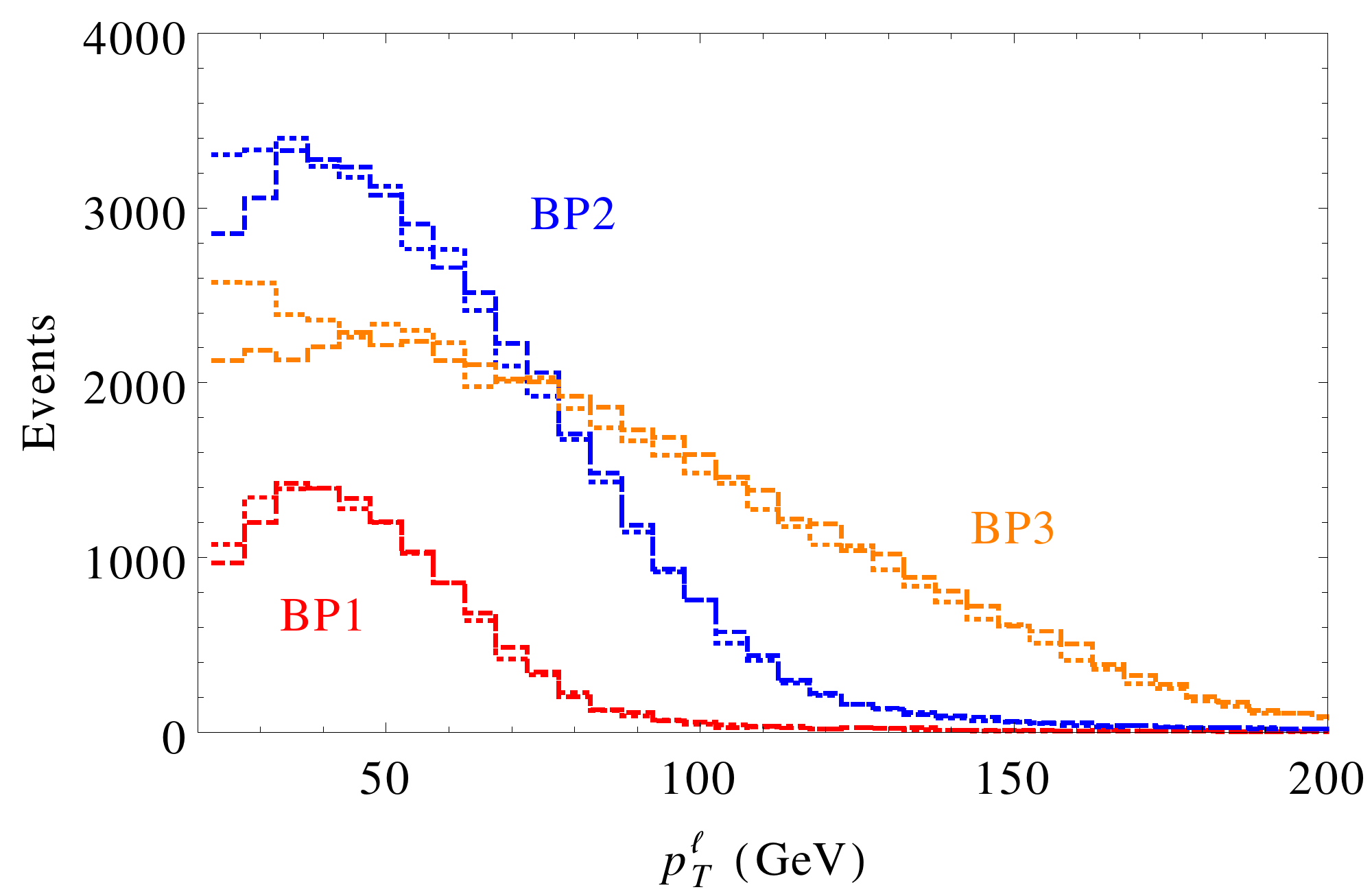}}}
\caption{The (a) lepton multiplicity and (b) lepton $p_T$ distribution for the benchmark points.}\label{lpf}
\end{center}
\end{figure}

\begin{figure}[thb]
\begin{center}
\hspace*{-2cm}

\includegraphics[width=0.6\linewidth]{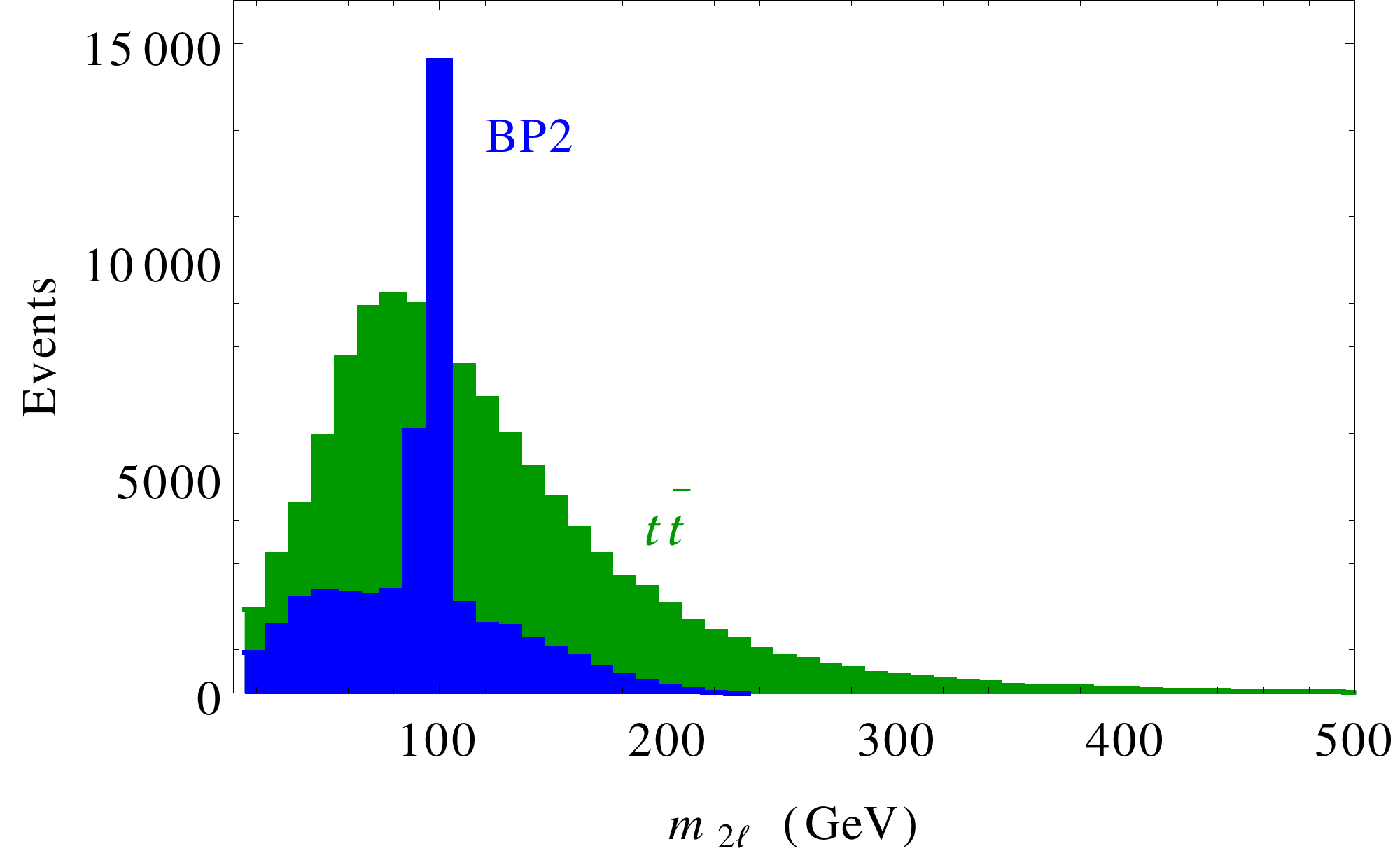}
\caption{The di-lepton invariant mass distribution for the signal BP2 and the background $t\bar{t}$.}\label{mll12}
\end{center}
\end{figure}

We have carried out a detailed analysis of the signal and of the background in a possible search for a light dilaton. For this purpose we have selected three benchmark points as given in Table~\ref{diltnbr}. The decay branching ratios given 
in Table~\ref{diltnbr} are independent of the conformal scale. For the benchmark point 1 (BP1), the dilaton is assumed to be of light mass of $200$ GeV, and its decay to the $H_{125}$ pair is not 
kinematically allowed. For this reason, as already mentioned, we look for slightly different final states in the analysis of such points. It appears evident that the dilaton may decay into gauge boson pairs when they are kinematically allowed. Such decays still remain dominant even after that the $t\bar{t}$ mode is open. This prompts us to study dilaton decays into $ZZ$, $WW$ via $3\ell$ and $4\ell$ final states. In the alternative case in which the dilaton also decays into a SM Higgs pair ($H_{125}$) along with gauge boson pairs, we have additional jets or leptons in the final states. This is due to the fact that 
the $H_{125}$ Higgs decays to the $WW$ and $ZZ$ pairs with one of the two gauge bosons off-shell (see Table~\ref{hbr}). We select two of such points when this occurs, denoted as BP2 and BP3, which are shown in Table~\ref{diltnbr}. Below we are going to present a separate analysis for each of the two cases. \\
 \begin{table}[t]
\begin{center}
\hspace*{-1.0cm}
\renewcommand{\arraystretch}{1.2}
\begin{tabular}{|c||c|c|c||}
\hline\hline
Decay&BP1 & BP2&BP3 \\
Modes &$m_\rho$ = 200 GeV&$m_\rho$ = 260 GeV&$m_\rho$ = 400 GeV\\
\hline
HH&-&0.245&0.290\\
\hline
$W^\pm W^\mp$&0.639&0.478&0.408\\
\hline
ZZ&0.227&0.205&0.191\\
\hline
$\tau\tau$&$2.54\times10^{-4}$&$7.8\times10^{-5}$&$2.05\times10^{-5}$\\
\hline
$\gamma \gamma$&$9.28\times10^{-5}$&$2.88\times10^{-5}$&$4.33\times10^{-6}$\\
\hline
$gg$&$0.131$&$0.0691$&$0.0390$\\
\hline
\hline
\end{tabular}
\caption{The benchmark points for a light dilaton with their mass-dependent decay branching ratios.}\label{diltnbr}
\end{center}
\end{table}
 \begin{table}
\begin{center}
\renewcommand{\arraystretch}{1.2}
\begin{tabular}{||c||c|c|c|c|c|c||}
\hline\hline
Decay Modes &$W^\pm W^\mp$&$Z\,Z$&$\bar b b$&$\bar \tau \tau$&$gg$&$\gamma\,\gamma$\\
\hline
$H_{125}$&0.208&0.0259&0.597&0.0630&0.0776&$2.30\times10^{-3}$\\
\hline
\hline
\end{tabular}
\caption{The corresponding branching ratios of the SM Higgs boson with a mass of 125 GeV.}\label{hbr}
\end{center}
\end{table}
The leptons in the final state are produced from the decays of the gauge bosons, which can come, in turn, either from the decay of the dilaton or from that of the $H_{125}$.  In such cases, for a dilaton sufficiently heavy, the four lepton signature ($4\ell$) of the final state is quite natural and their momentum configuration will be boosted. In Fig.~\ref{lpf}(a) we show the multiplicity distribution of the leptons and  in Fig.~\ref{lpf}(b) their $p_T$ distribution for the chosen benchmark points. Here the lepton multiplicity has been subjected to some basic cuts on their transverse momenta ($p_T\geq 20$) GeV and isolation criteria given earlier in this section. Thus soft and non-isolated leptons are automatically cut out from the distribution. From Fig.~\ref{lpf}(b) it is clear that the leptons in BP3 can have a very hard transverse momentum ($p_T \sim 200$ GeV), as the corresponding dilaton is of $400$ GeV. Notice that the di-lepton invariant mass distribution in Fig.~\ref{mll12} presents a mass peak around $m_Z$ for the signal (BP2) but not for the dominant SM top/antitop ($t\bar{t}$) background. This will be used later as a potential selection cut in order to reduce some of the SM backgrounds.

\subsection{Light dilaton: $m_{\rho} < 2m_{H_{125}}$}
 In this subsection we analyse final states with at least three  ($\geq 3\ell +X +\etmiss$) and 4 ($\geq 4\ell +X +\etmiss$) leptons (inclusive) and missing transverse energy that can result from the decays of the dilaton into $ZZ$, where we consider the potential  SM backgrounds. The reason for considering the $3\ell$ final states is because one of the four leptons ($4\ell$) could be missed. This is in general possible due to the presence of additional kinematical cuts introduced when hadronic final states are accompanied by leptons. We present a list of the number of events 
 for the $3\ell$ and $4\ell$ final states in Table~\ref{BP1n} for BP1, and the dominant SM backgrounds at integrated luminosity of 100 fb$^{-1}$ at the LHC. The potential SM backgrounds come from the $t\bar{t}Z$ and  $tZW$ sectors, from intermediate gauge boson pairs ($VV$) and from the triple gauge boson vertices $VVV$ ($V: W^\pm, Z$). Due to the large $t\bar{t}$ cross-section, with the third and fourth lepton - which can originate from the corresponding $b$ decays -  this background appears to be an irreducible one. For this reason we are going to apply successive cuts for its further reduction, as described in Table~\ref{BP1n}.
 \begin{table}[t]
\begin{center}
\hspace*{-1.0cm}
\renewcommand{\arraystretch}{1.3}
\begin{tabular}{|c||c||c|c|c|c|c||}
\hline\hline
Final states&\multicolumn{1}{|c||}{Benchmark}&\multicolumn{5}{|c||}{Backgrounds }
\\
\hline
&BP1  & $t\bar{t}$& $t\bar{t}Z$ &$tZW$&$VV$& $VVV$\\
\hline
\hline
$\geq 3\ell \,+\, \ptmiss \leq 30\, \rm{GeV}$&494.97&275.52&65.17&22.29&6879.42&765.11\\
$\,+\,|m_{ll}-m_Z|<5\,\rm{GeV}$&384.47&68.88&62.68&20.93&2514.92&16.16\\
$\,+\,n_{\rm{b_{jet}}}=0$&377.56&9.84&17.64&10.08&2479.66&15.13\\
\hline
Significance&7.00&\multicolumn{5}{|c||}{}\\
\hline
$\mathcal L_5$&51 fb$^{-1}$&\multicolumn{5}{|c||}{}\\
\hline
\hline
$\geq 4\ell \,+\, \ptmiss \leq 30\, \rm{GeV}$&273.96&0.00&3.32&1.36&1655.99&34.18\\
$\,+\,|m_{ll}-m_Z|<5\,\rm{GeV}$&218.71&0.00&3.11&1.16&627.38&4.44\\
\hline
Significance&7.48&\multicolumn{5}{|c||}{}\\
\hline
$\mathcal L_5$&45 fb$^{-1}$&\multicolumn{5}{|c||}{}\\
\hline
\hline
\end{tabular}
\caption{Numbers of events for the $3\ell+\ptmiss$  and $4\ell$ final states for the BP1 and 
the dominant SM backgrounds, at an integrated luminosity of $100$ fb $^{-1}$.}\label{BP1n}
\end{center}
\end{table}

The primary signal that is considered is characterised  by the kinematical cut $3\ell +\ptmiss \leq 30$ GeV. The choice of a very low missing $p_T$ is justified because when both $Z$'s decay to charged lepton pairs they give rise to $\geq 3\ell$ and $\geq 4\ell$ final states which are neutrinoless. The theoretical prediction of no missing energy, however, cannot be fully satisfied as the missing transverse momentum $\ptmiss$ is calculated by estimating the total visible $p_T$ of the jets and of the leptons after the threshold cuts. Next we demand the di-lepton be characterised by an invariant mass around $Z$ mass i.e., $\,|m_{ll}-m_Z|<5\,\rm{GeV}$, which reduces the $t\bar{t}$, 
$VV$ and $VVV$ backgrounds quite significantly. A further requirement of no $b$-jet ( i.e., $n_b=0$) reduces the $t\bar{t}, $$t\bar{t}Z$ and $tZW$ backgrounds. By looking at the signal, we observe that these cuts do not affect the signal number for BP1. 
After imposing all the cuts, we find that an integrated luminosity of $\mathcal{O}(51)$ fb$^{-1}$ is required for 
a $5\sigma$ reach in this final state.  The demand of $4\ell$ of course reduces the background but also reduces the signal event numbers. In this case  $\mathcal{O}(45)$ fb$^{-1}$ of integrated luminosity 
is required for a $5\sigma$ discovery.

 \subsection{Heavy dilaton: $m_{\rho} > 2m_{H_{125}}$}
  In this case we consider points where $m_{\rho} > 2m_{H_{125}}$, allowing decays of the dilaton to $H_{125}$ pairs. For this purpose we have chosen two benchmark points, one with $m_{\rho}=260$ GeV - where the channel $\rho \to\, H_{125}H_{125}$ is just open - and another one with  $m_{\rho}=400$ GeV, where even the $\rho \to t\bar{t}$ channel is open.  The decay mode via a $H_{125}$ pair, in turn decaying into gauge boson pairs, gives additional jets which accompany the $3\ell$ and $4\ell$ final states and help in a further reduction of the SM backgrounds.
  
Table~\ref{4l}  presents the number of expected events generated at the BP2  and BP3 benchmark points for the signal and for the dominant SM backgrounds. Here we have considered $\geq 3\ell$ GeV and $\geq 4\ell$ final states respectively, at an integrated luminosity of $1000$ fb$^{-1}$. The dominant backgrounds are as before, and listed in Table~\ref{4l}. Notice that if we demand the tagging of at least two additional jets and the $b$-jet veto, we can reduce the backgrounds even further. The result shows that in the case of BP2 and BP3 a dilaton signal could be discovered at an integrated luminosity of $\mathcal{O}(130)$ and $\mathcal{O}(570)$ fb$^{-1}$ respectively for the $\geq 3\ell$ final state. For the  $\geq 4\ell$  f a $5\sigma$ discovery reach can be achieved even with 114 fb$^{-1}$ and 374 fb$^{-1}$ of integrated luminosity for BP2 and BP3 respectively.

 \begin{table}[t]
\begin{center}
\hspace*{-1.0cm}
\renewcommand{\arraystretch}{1.3}
\begin{tabular}{|c||c|c||c|c|c|c|c||}
\hline\hline
Final states&\multicolumn{2}{|c||}{Benchmark}&\multicolumn{5}{|c||}{Backgrounds }
\\
\hline
& BP2&BP3 & $t\bar{t}$& $t\bar{t}Z$  &$tZW$&$VV$& $VVV$\\
\hline
\hline
$\geq 3\ell$&3882.08&1642.28&10725.9&4790.19&1364.73&177140&53660.2\\
$\,+\, n_{\rm{b_{jet}}}=0$&3812.82&1627.53&5510.54&1550.38&664.92&176167&53604.8\\
$\,+\,n_{\rm{jet}}\geq2$&2677.82&1255.06&2952.08&1469.43&579.62&29165.5&324.28\\
\hline
Significance&13.89&6.64&\multicolumn{5}{|c||}{}\\
\hline
$\mathcal L_5$&130 fb$^{-1}$&568 fb$^{-1}$&\multicolumn{5}{|c||}{}\\
\hline
\hline
$\geq 4\ell$&1400.47&678.55&0.00&502.26&149.27&17338.1&2379.06\\
$\,+\,n_{\rm{jet}}\geq2\,+\, n_{\rm{b_{jet}}}=0$&865.68&448.68&0.00&147.36&48.46&2334.44&36.13\\
\hline
Significance&14.78&8.17&\multicolumn{5}{|c||}{}\\
\hline
$\mathcal L_5$&114 fb$^{-1}$&374 fb$^{-1}$&\multicolumn{5}{|c||}{}\\
\hline
\hline
\end{tabular}
\caption{We present the final state numbers for $4\ell+\ptmiss$ final states for the benchmark points and 
the dominant SM backgrounds at an integrated luminosity of $1000$ fb $^{-1}$.}\label{4l}
\end{center}
\end{table}

\begin{figure}[thb]
\begin{center}
\hspace*{-2cm}
\mbox{\subfigure[]{
\includegraphics[width=0.59\linewidth]{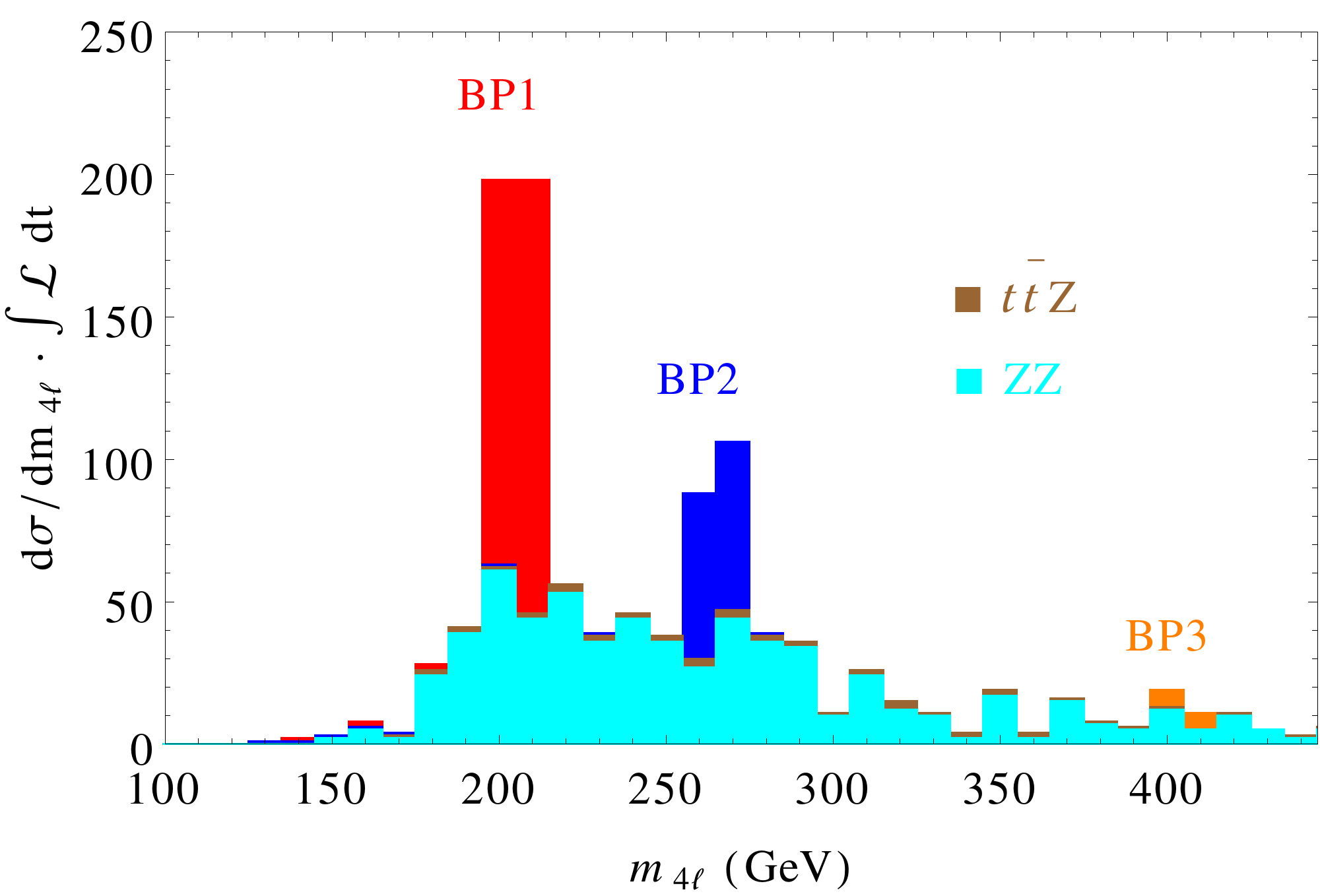}}
\hspace*{.5cm}
\subfigure[]{\includegraphics[width=0.62\linewidth]{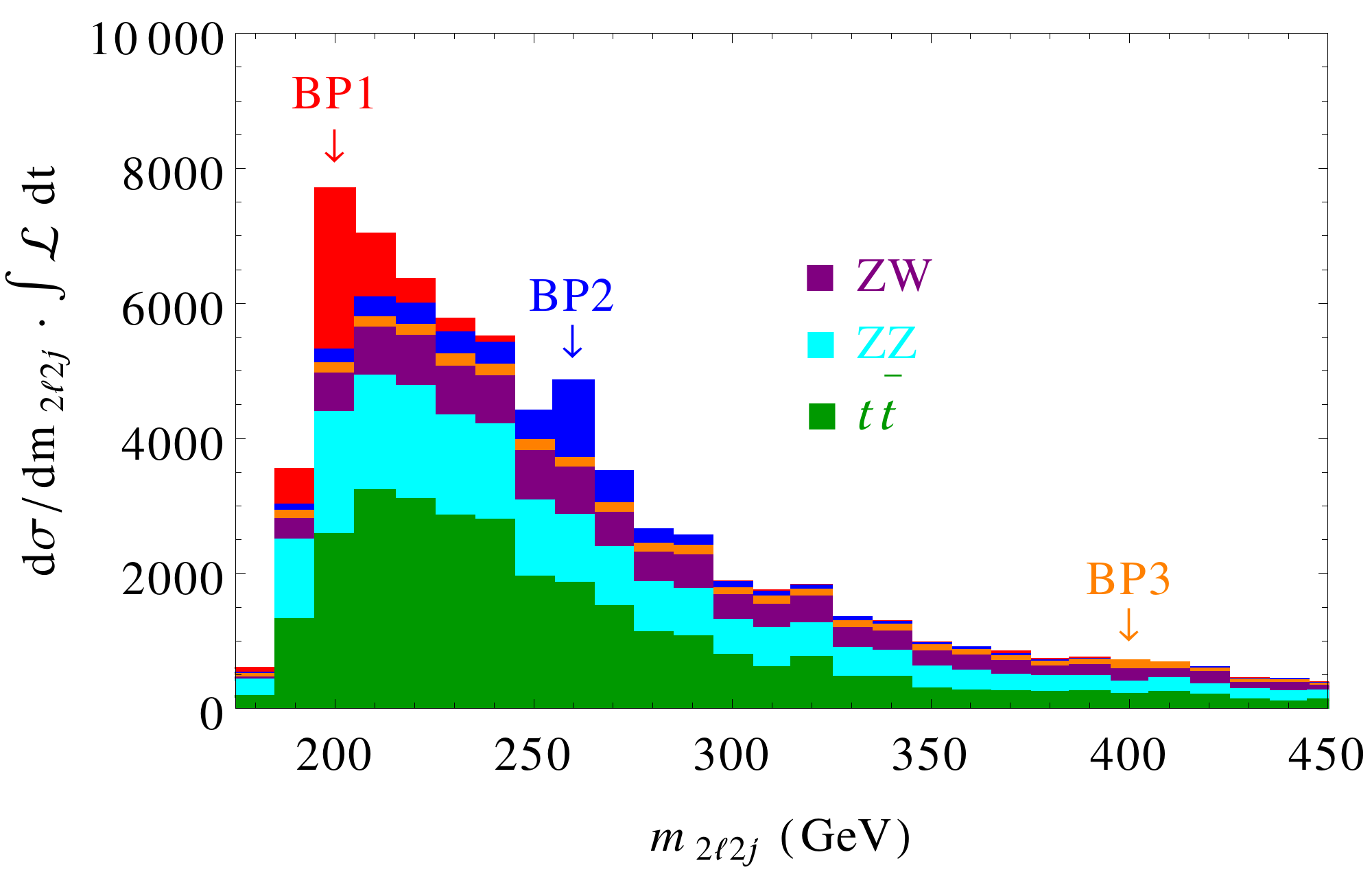}}}
\caption{The invariant mass distribution for the benchmark points and the dominant SM backgrounds 
for $4\ell$ and $2\ell2j$ final state respectively at an integrated luminosity of 100 fb$^{-1}$.}\label{invmass}
\end{center}
\end{figure}
Next we try to reconstruct the dilaton mass peak from the $\geq 4\ell$ and $2\ell\, 2j$ channels. 
In the first case we consider the isolated $4\ell$'s after enforcing the basic cuts, and then demand that the di-leptons are coming from the $Z$ boson mass peak. This guarantees that we are reconstructing either the $\rho\to ZZ$ or the $\rho \to H_{125}H_{125}\to ZZ+X$ incoming channel. Fig.~\ref{invmass}(a) shows the plot of the invariant 
mass distributions $m_{4\ell}$ for all three benchmark points, along with the dominant backgrounds. The presence of a clear mass peak certainly allows the reconstruction of the dilaton mass. We have selected the number of events around the mass peaks, i.e., $|m_{4\ell} -m_{\rho}|\leq 10$ GeV for the benchmark points, which are shown in Table~\ref{4lpeak} at an integrated luminosity of 100 fb$^{-1}$. It is clear that for the BP1 and BP2 benchmark points the mass peak can be resolved with very early data at the LHC, with a 14 TeV run.\\
Fig.~\ref{invmass}(b) shows the invariant mass distribution, where we consider a pair of charged leptons around the $Z$ mass peak, i.e., $|m_{\ell\ell}-m_Z|<5\,\rm{GeV}$ as well as a pair of jets, i.e., $|m_{jj}-m_Z|< 10\,\rm{GeV}$. Such di-jet pairs and di-lepton pairs are then taken in all possible combinatorics to evaluate the $m_{\ell\ell jj}$ mass distribution, as shown in Fig.~\ref{invmass}(b). Clearly the $Y$ axis of the figure shows such possible pairings and the $X$ axis indicates the mass scale. We see the right combinations peak, which sits around the benchmark points. We have also taken the dominant backgrounds with their combinatorics to reproduce the invariant mass $m_{\ell\ell jj}$. In Table~\ref{2l2jlpeak}
we list the results around the mass peak, i.e. for $|m_{2\ell2j} -m_{\rho}|\leq 10$ GeV. It is easily observed that such constraint can be a very handy guide to identify the resonance mass peak using very early data at the LHC with 14 TeV.

 \begin{table}[t]
\begin{center}
\hspace*{-1.0cm}
\renewcommand{\arraystretch}{1.5}
\begin{tabular}{c||c|c|c||}
\hhline{~===}
&\multicolumn{3}{c||}{Number of events in}\\
&\multicolumn{3}{c||}{$|m_{4\ell} -m_{\rho}|\leq 10$ GeV}\\
\hhline{~===}
& BP1&BP2&BP3\\
\hline
\hline
\multicolumn{1}{||c||}{Signal}&396&194&30\\
\hline
\multicolumn{1}{||c||}{Background}&108&77&18\\
\hline
\multicolumn{1}{||c||}{Significance}&17.64&11.78&4.33\\
\hline
\hline
\end{tabular}
\caption{We present the events number for $\geq 4\ell$ final state around the dilaton mass peak, i.e. $|m_{4\ell} -m_{\rho}|\leq 10$ GeV, for the benchmark points and the backgrounds at an integrated luminosity of 100 fb$^{-1}$.}\label{4lpeak}
\end{center}
\end{table}

 \begin{table}[t]
\begin{center}
\hspace*{-1.0cm}
\renewcommand{\arraystretch}{1.5}
\begin{tabular}{c||c|c|c||}
\hhline{~===}
&\multicolumn{3}{c||}{Number of events in}\\
&\multicolumn{3}{c||}{$|m_{\ell\ell jj} -m_{\rho}|\leq 10$ GeV}\\
\hhline{~===}
& BP1&BP2&BP3\\
\hline
\hline
\multicolumn{1}{||c||}{Signal}&14727&8371&1390\\
\hline
\multicolumn{1}{||c||}{Background}&10887&6706&1234\\
\hline
\multicolumn{1}{||c||}{Significance}&92.02&68.17&27.13\\
\hline
\hline
\end{tabular}
\caption{We present the events number for $\geq 2\ell$ final state around the dilaton mass peak, i.e. $|m_{2\ell2j} -m_{\rho}|\leq 10$ GeV, for BP1, BP2, BP3 and the backgrounds  at an integrated luminosity of 100 fb$^{-1}$.}\label{2l2jlpeak}
\end{center}
\end{table}

\section{Perspectives on compositeness and $\xi$ dependence }
\label{non0xi}
\begin{figure}[thb]
\begin{center}
\hspace*{-2.5cm}
\mbox{\subfigure[]{
\includegraphics[width=0.62\linewidth]{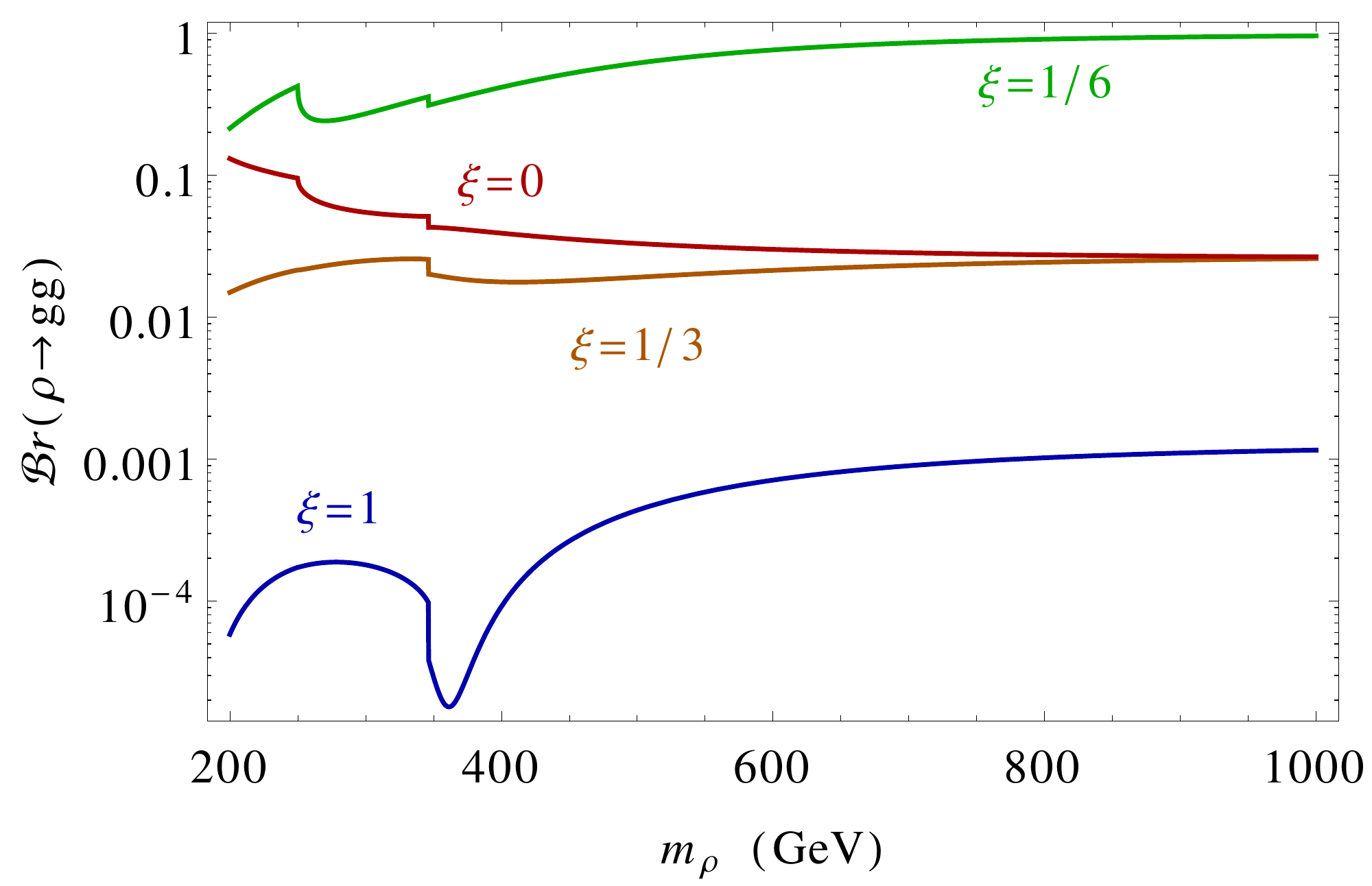}}
\hspace*{0.3cm}
\subfigure[]{\includegraphics[width=0.62\linewidth]{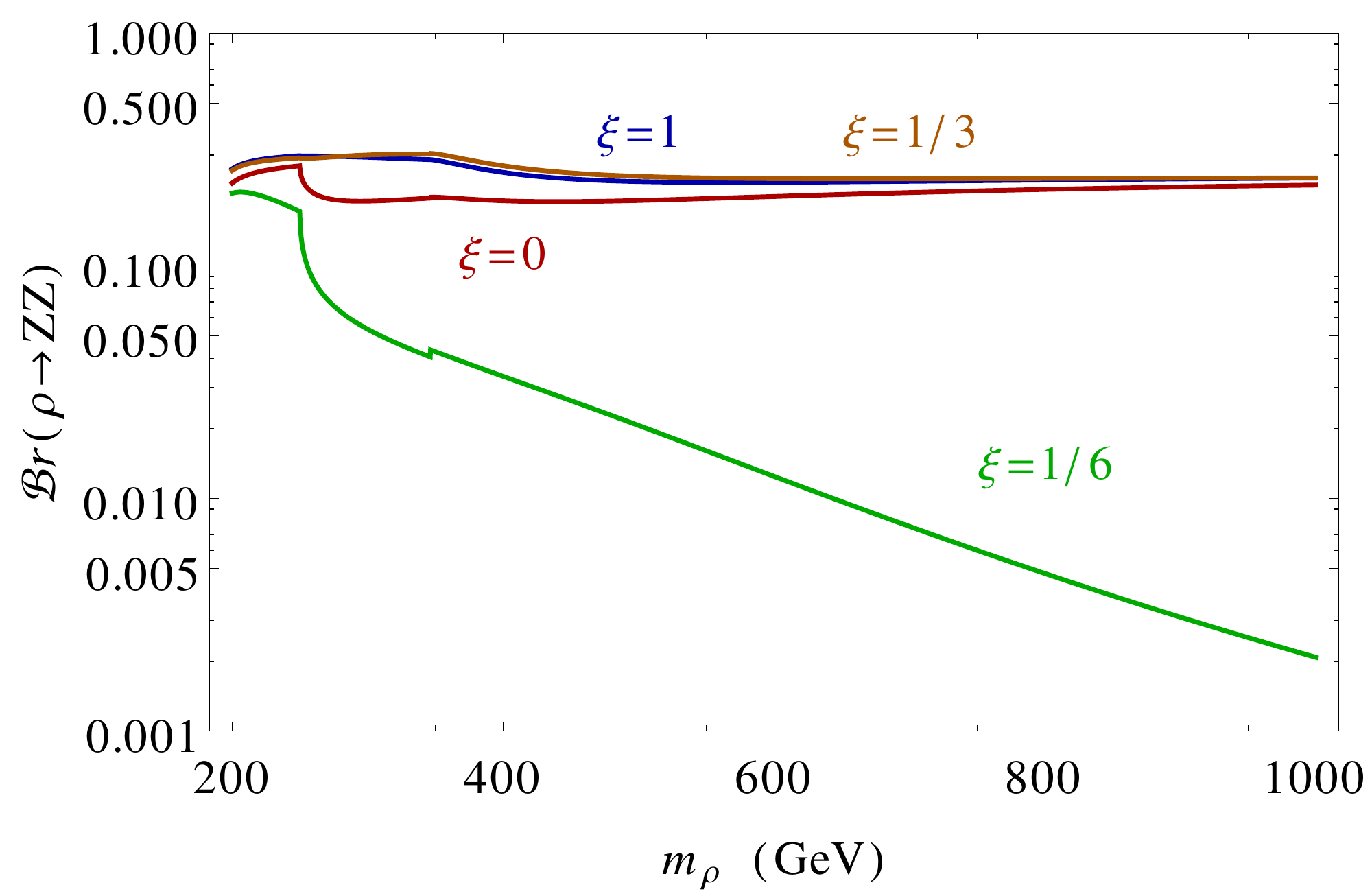}}}
\hspace*{-2.5cm}
\mbox{\subfigure[]{
\includegraphics[width=0.62\linewidth]{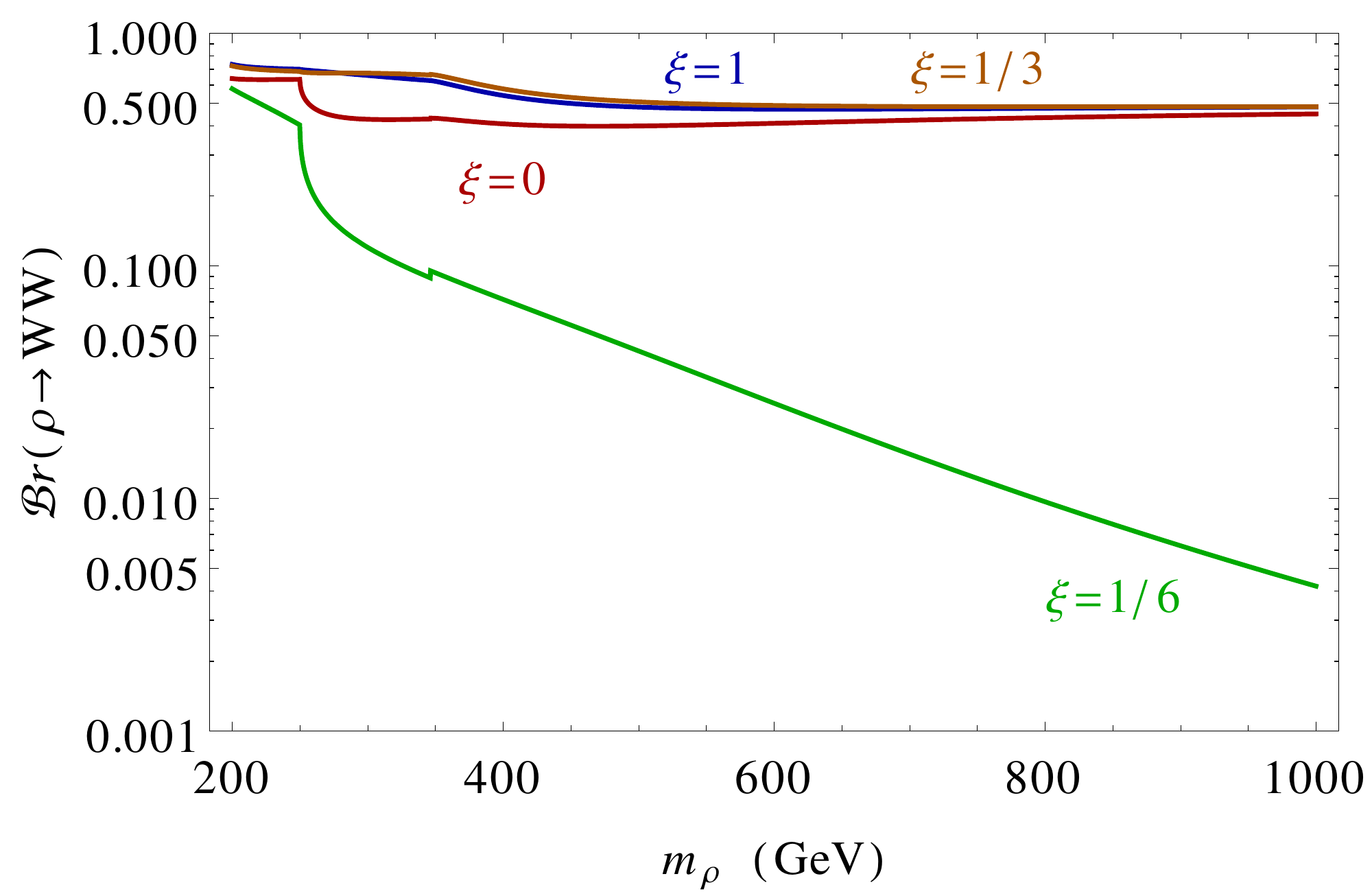}} \hspace{0.3cm}   
\subfigure[]{\includegraphics[width=0.62\linewidth]{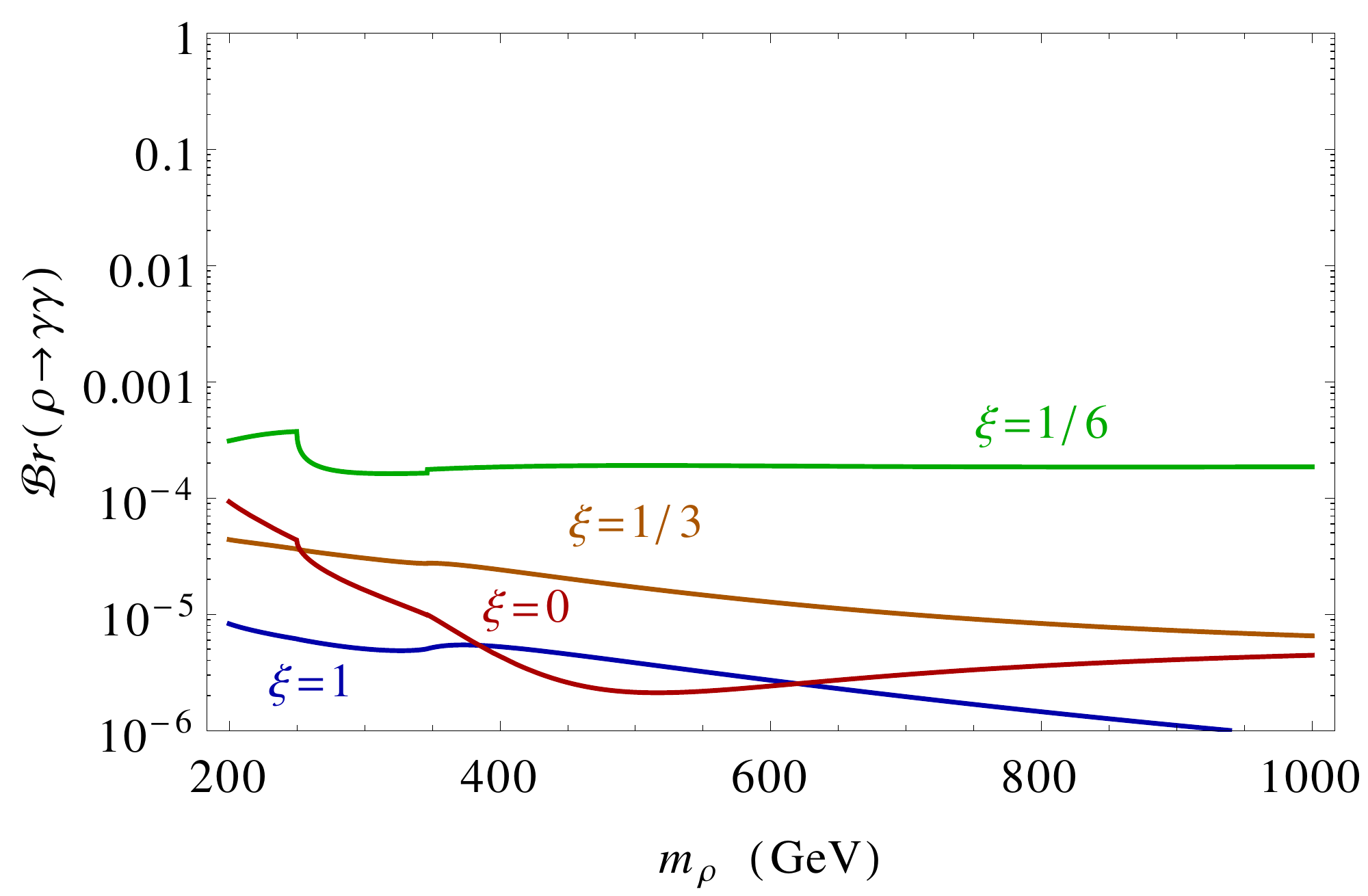}}}
\caption{The decay branching ratios of the dilaton (a) to gluons, (b)-(c) massive gauge bosons 
and (d) photons pairs for different $\xi$ parameters.}\label{diffxi}
\end{center}
\end{figure}

\begin{figure}[thb]
\begin{center}
\hspace*{-2.3cm}
\mbox{\subfigure[]{
\includegraphics[width=0.62\linewidth]{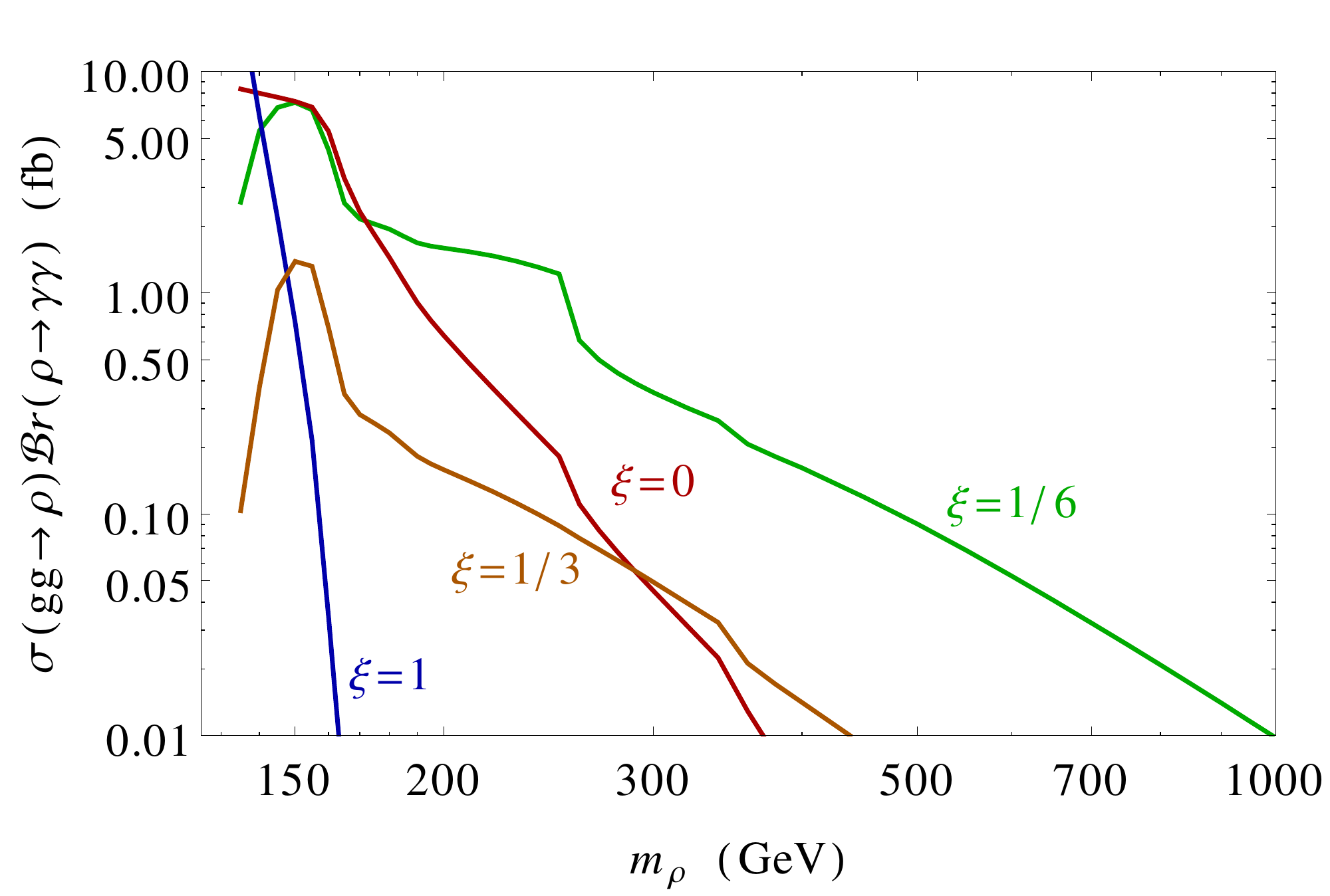}}
\hspace*{0.3cm}
\subfigure[]{
\includegraphics[width=0.62\linewidth]{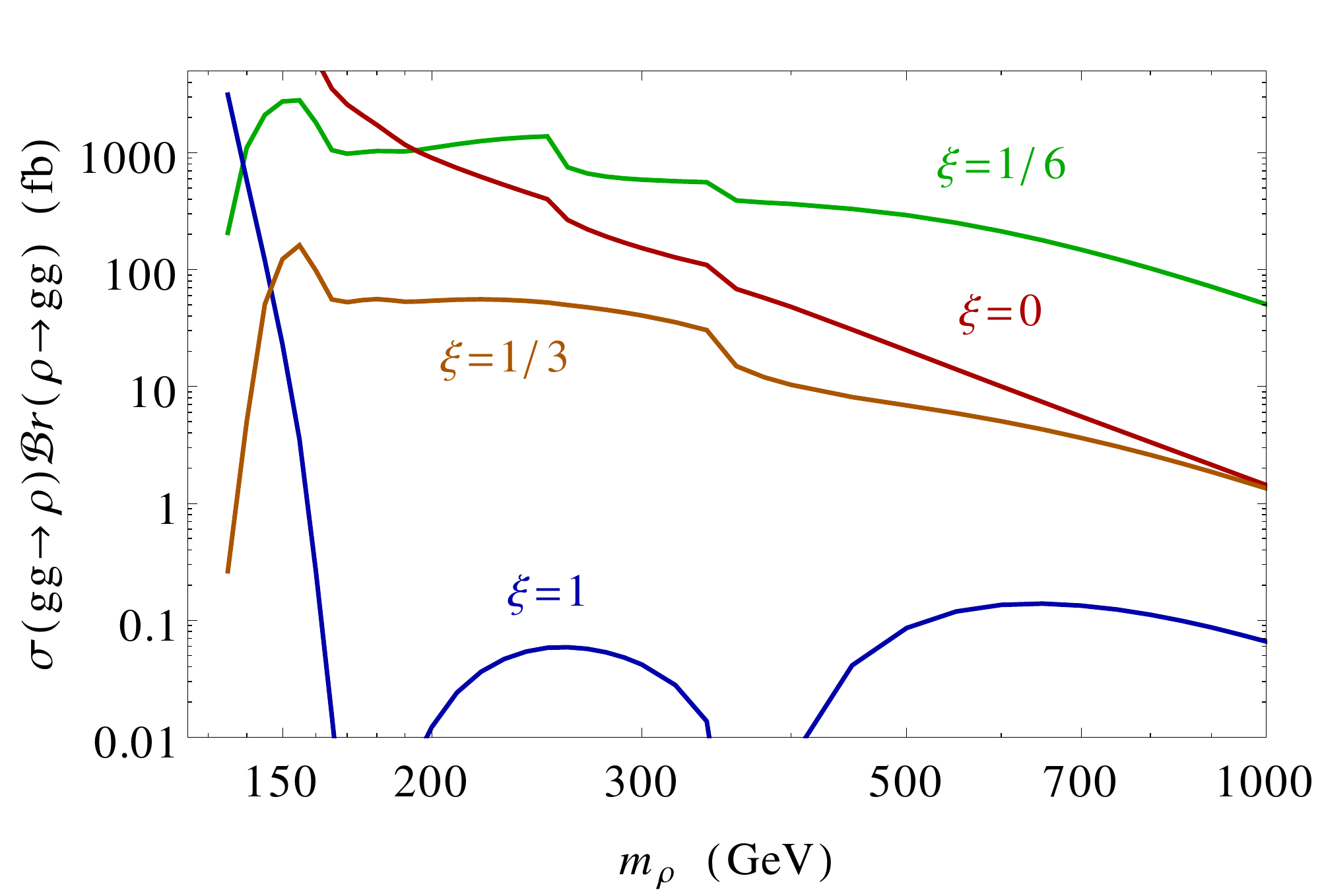}}}
\caption{Di-gluon and di-photon signal of a dilaton for a varying $\xi$.}\label{xsgldi}
\end{center}
\end{figure}

In our analysis the dilaton has been treated as a fundamental state, with interactions which are dictated from 
Eq. (\ref{tmunu}). The perturbative analysis that follows from this interaction does not take into account possible effects of compositeness, which would involve the wave function of this state both in its production and decay. In this respect, this treatment is quite similar to the study of the $\pi\to \gamma\gamma$ decay using only the divergence of the interpolating axial-vector current rather then the pion itself, with its hadronic wave function now replaced by the divergence of the dilatation current $J_D$. 
Those effects could modify the predictions that emerge from our analysis. \\
Another possible modification of our results will be certainly linked to a nonzero value of the $\xi$ parameter. The search for a valuable signal of a nonminimal dilaton at the LHC requires a completely independent calibration of the kinematical cuts that we have discussed. While we hope to address this point in a future work, we can however obtain a glimpse of the dependence of the signal (production/decays) as a function of $\xi$. \\
This behaviour is clearly illustrated in Fig.~\ref{diffxi} where the decay into massless and massive states of a conformal dilaton are  dependent on the  improvement coefficient  $\xi$. Fig.~\ref{diffxi}(a), (d)  show the decay branching fraction to gluon and photon pair respectively.  We see that for $\xi=1/6$ they are enhanced compared to other values of $\xi$. Similarly, the massive gauge bosons modes are suppressed for $\xi=1/6$ as can be seen from  Fig.~\ref{diffxi}(b),(c).  In Fig.~\ref{xsgldi}
we present the production cross-sections for di-gluons and di-photon final states. Notice that for 
$\xi=1/6$ these two modes have much larger rates than for other $\xi$ cases.  Unlike the minimal case of $\xi=0$, the $\xi=1/6$ can be studied via di-jet or di-photon final states. 

It is expected that a dilaton which arises from the breaking of a conformal symmetry should be described by a conformal coupling $\xi=1/6$, at least in the high energy limit. The signature of such a state, if composite, is in the anomaly pole of correlators involving the dilatation current and two vector currents, as pointed out in \cite{CDS}. The dilatation current inherits the same pole from the $TVV$ correlator \cite{Giannotti:2008cv,ACD2,ACD0} while
the explicit/non perturbative breaking of the conformal symmetry would then be responsible for the generation of its mass. \\
In a more general framework, the possibility of having similar states in superconformal theories has been extensively discussed in \cite{CCDS} from a perturbative side. It has been shown, for instance, that classical superconformal theories are characterised by a complete alignment in their conformal anomaly multiplets. An axion/dilaton/dilatino composite multiplet would then be the natural manifestation of this alignment found in the superconformal anomaly action.

\section{Conclusions}\label{concl}
 In this article we have performed a study of a minimally coupled ($\xi=0$) dilaton, identifying signature for its detection at the LHC via multi-leptonic final state. A detailed signal vs background analysis shows that very early data at the LHC $\mathcal{O}(50)$ fb$^{-1}$ can probe some of the benchmark points that we have selected, as examples valid beyond the $5\sigma$ significance. A dilaton with a mass of about 400 GeV can be probed with $\mathcal{O}(400)$ fb$^{-1}$ of integrated luminosity. $4\ell$ and $\ell\ell jj$ multi-lepton final states provide a significant channel for the discovery of such a resonance peak. We have shown that current data at 7 and 8 TeV do not exclude a conformal scale of 5 TeV. A conformally coupled dilaton, in particular, is characterised by larger production and decay rates into massless vector channels, offering a signal which could be of specific interest for current and future analysis at the LHC. The results of our study can be extended by considering higher conformal breaking scales $\Lambda$ and for a nonminimally coupled dilaton. In the case of a nonminimal coupling the search can be performed using di-photon and di-jet final states rather than the multi-lepton channel that we have discussed above. The details of such investigation need to take care of different set of SM backgrounds, cuts, and so on, which require a separate analysis, that we leave to a future work. 
 
\section*{Acknowledgments}
L.D.R. is supported by the "Angelo Della Riccia'' foundation. The work of C.C. is
partially supported by a The Leverhulme Trust Visiting Professorship at the University of Southampton in
the STAG Research Centre and Mathematical Sciences Department, where
part of his work has been carried out.
  

\end{document}